\documentclass[aps,pra,twocolumn,groupedaddress,showpacs,showkeys,floatfix]{revtex4-1}

\usepackage{amsmath,amssymb}
\usepackage[pdftex]{graphicx}
\usepackage{ascmac}
\usepackage{amsthm}
\usepackage{braket}
\usepackage{bm}
\usepackage{mathrsfs}
\usepackage{makeidx}
\usepackage{txfonts}
\usepackage{endnotes}

\begin{document}


\title{Two-Dimensional Thouless Pumping of Ultracold Fermions in Obliquely Introduced Optical Superlattice}


\author{Fuyuki Matsuda}
\email[]{fuyuki.matsuda@scphys.kyoto-u.ac.jp}
\affiliation{Department of Physics, Kyoto University, Kyoto 606-8502, Japan}

\author{Masaki Tezuka}
\email[]{tezuka@scphys.kyoto-u.ac.jp}
\affiliation{Department of Physics, Kyoto University, Kyoto 606-8502, Japan}

\author{Norio Kawakami}
\email[]{norio@scphys.kyoto-u.ac.jp}
\affiliation{Department of Physics, Kyoto University, Kyoto 606-8502, Japan}



\begin{abstract}
We propose a two-dimensional (2D) version of Thouless pumping that can be realized by using ultracold atoms in optical lattices. To be specific, we consider a 2D square lattice tight-binding model with an obliquely introduced superlattice. It is demonstrated that quantized particle transport occurs in this system, and that the transport is expressed as a solution of a Diophantine equation. This topological nature can be understood by mapping the Hamiltonian to a three-dimensional (3D) cubic lattice model with a homogeneous magnetic field. We also propose a continuum model with obliquely introduced superlattice and obtain the amount of pumping by calculating the Berry curvature. For this model, the same Diophantine equation can be derived from the plane-wave approximation. Furthermore, we investigate the effect of a harmonic trap by solving the time-dependent Schr\"odinger equation. 
Under a harmonic trap potential, as often used in cold atom experiments, we show, by numerical simulations, that nearly quantized pumping occurs when the phase of the superlattice potential is driven at a moderate speed. Also, we find that two regions appear, the Hofstadter region and the rectifying region, depending on the modulation amplitude of the superlattice potential. In the rectifying region with larger modulation amplitudes, we uncover that the pumping direction is restricted to exactly the $x$-axis or the $y$-axis direction. This difference in these two regions causes a crossover behavior, characterizing the effect of the harmonic trap.
\end{abstract}
\pacs{}
\maketitle

\section{Introduction}
Intensive studies of the quantum Hall effect have opened up a wide research area known as topological phases \cite{RevModPhys.82.3045, RevModPhys.83.1057}. Topological phases are characterized by the associated invariants, which are prohibited from varying unless the energy gap closes. Moreover, topological phases have a characteristic property called the bulk-edge correspondence\cite{PhysRevB.23.5632, PhysRevB.25.2185, PhysRevLett.71.3697, RevModPhys.83.1057}, which guarantees the existence of the edge states, meaning that electrons can only move on the edge of a given material, as long as the bulk topological invariant has a nontrivial value. In quantum Hall systems, the topological invariant is defined by the Chern number\cite{PhysRevLett.49.405, PhysRevLett.51.51, PhysRevB.31.3372, KOHMOTO1985343}, and it appears as the quantized Hall conductance. Similar relationships also appear in three-dimensional (3D) quantum Hall systems\cite{PhysRevB.41.11417,PhysRevB.45.13488, Koshino2001, Koshino2002}.

Topological phases can be realized in various experimental platforms, not only in materials, but also in, for example, optical waveguides\cite{Kraus2012} and photonic systems\cite{PhysRevA.84.043804, NaturePhys.7.907, NatureMat.12.233}. In particular, a great amount of effort has been made to realize topological insulators in cold atom systems by using synthetic gauge fields, which are induced by an optical lattice\cite{Dalibard2011}. Various topological systems have been realized in cold atom systems, for instance, the Su--Schriefer--Heeger model \cite{Nature.9.795}, the Hofstadter model \cite{PhysRevLett.111.185301, PhysRevLett.111.185302}, synthetic dimensions \cite{Mancini1510, Stuhl1514}, and the Haldane model \cite{Nature.515.237}. Recently, topological charge pumping, called Thouless pumping\cite{Review1983, PhysRevB.34.5093}, has been realized in cold atom systems \cite{NaturePhys.12.296, NaturePhys.12.350}, and many theoretical proposals have been made in numerous studies\cite{PhysRevA.57.R2278, PhysRevA.76.052304, PhysRevA.84.013608, PhysRevLett.111.026802, PhysRevA.90.063638, PhysRevA.92.013612, PhysRevA.92.013609, Marra2017,PhysRevB.98.165128, JPSJ.83.083707,marra2020topologically,PhysRevB.98.115147,PhysRevLett.118.230402, PhysRevA.95.063630, PhysRevB.98.245148, PhysRevLett.120.106601, PhysRevB.100.064302, PhysRevA.101.053630, Arceci_2020}. Also, two-dimensional (2D) Thouless pumping, where the nonlinear response is characterized by a four-dimensional (4D) topological invariant, is reported in Ref. \onlinecite{Nature.553.55}.

Mapping the Hamiltonian between different dimensions plays an important role in understanding topological properties. One of the significant examples of mapping can be seen in the problem of the Hofstadter butterfly\cite{PhysRevB.14.2239}. Hofstadter mapped a 2D system of square lattice under a uniform magnetic field to a one-dimensional (1D) equation, which is called the Harper equation. Interestingly, in a recent study of Thouless pumping, this mapping was used backwards, and the topological equivalence over a certain group of Hamiltonians was shown. This kind of mapping to higher dimensions can be applied not only to 1D systems, but also to 2D and higher dimensional systems\cite{Kraus2013}. 

In this study, we investigate topological properties for a 2D tight-binding model of particles that hop on a square lattice in the presence of modulation in the on-site potential, where the modulation amplitude depends only on the distance from an inclined line. We find that this model has a topological nature that can be understood clearly by applying a mapping from this 2D system to the corresponding 3D system. 
We show that the amount of pumping after the phase of the superlattice is adiabatically changed by one cycle is represented by a set of Chern numbers, which correspond to the solution of a Diophantine equation. 

Furthermore, we consider a continuum model to confirm the result of the tight-binding model, and also to take the multiband effect fully into account. In the continuum model, as have been employed in Ref. \onlinecite{Wang2013}, the contribution of the higher bands is included, while it is ignored in the tight-binding model. We discuss Thouless pumping by using the plane-wave approximation as a counterpart of the tight-binding model. 
Moreover, we investigate how the harmonic trap affects our model by solving the time-dependent Schr\"{o}dinger equation. 
We show that nearly quantized pumping occurs when the phase of the superimposed potential is changed at an appropriate speed. We find that the pumping behavior is different between two regions: In the Hofstadter region with smaller modulation amplitudes, the amount of pumping obeys the Diophantine equation, while in the rectifying region with larger modulation amplitudes, the pumping direction is restricted to exactly the $x$-axis or the $y$-axis direction. 
This difference causes a crossover behavior, highlighting the effect of the harmonic trap.

This paper is organized as follows. In sect. 2, we introduce a 2D square lattice tight-binding model with an obliquely introduced potential in order to investigate the effect of the superimposed potential in the context of topological phases. 
In sect. 3, we study a continuum model that corresponds to the above-mentioned tight-binding model. 
In sect. 4, we conclude our paper.

\section{Topological Aspects of 2D Optical Lattice with Obliquely Introduced Potential}
\subsection{Model}
Recently, the development of experimental techniques regarding cold atoms in an optical lattice has been remarkable\cite{PhysRevA.69.013603, RevModPhys.80.885}, and the degree of freedom in designing a quantum simulation is high. 
Triangular\cite{Becker_2010}, square\cite{Yang}, and hexagonal lattices\cite{nphys1916} have all been realized in cold atom systems.
Moreover, it is possible to realize nonstandard lattices in cold atom systems. One example of such lattices is the lattice made by superimposing a typical lattice potential and a superlattice potential. 
For example, in Ref. \onlinecite{Taiee1500854}, Taie {\it et al.} realized an optical Lieb lattice by adding a diagonal lattice to a conventional square optical lattice. By taking advantage of these high degrees of freedom of the optical lattice, various topological phases have been proposed theoretically\cite{PhysRevLett.100.070402, PhysRevA.79.053639, PhysRevB.84.165115, PhysRevA.86.031603, PhysRevA.82.013608, PhysRevA.93.043617, PhysRevA.94.043611, PhysRevA.96.032103, PhysRevLett.121.133002} and realized experimentally\cite{Goldman6736, ncomms2523, doi:10.1063/1.5087957, NatPhys.15.911, deLeseleuc775, RevModPhys.91.015005}. In particular, the cold atom system is one of the most suitable platforms for realizing Thouless pumping. Various types of Thouless pumping have been realized in cold atom experiments\cite{NaturePhys.12.296, NaturePhys.12.350, PhysRevLett.117.170405, Nature.553.55}.

Also, this study is partly motivated by recent remarkable progress in solid-state physics, i.e. twisted bilayer graphene (TBG). TBG has been receiving much attention lately because of the realization of unconventional superconductivity\cite{Nature26160}. TBG has been investigated in many contexts, such as its electronic structures\cite{PhysRevLett.99.256802, ROZHKOV20161}, flat bands\cite{PhysRevB.82.121407}, topological band structures\cite{PhysRevB.84.045436}, and Moir\'e butterflies\cite{PhysRevB.84.035440}. It also provides a prototypical example of superimposed lattice potential.

\begin{figure}[t]
\begin{center}
\includegraphics[width=5.5cm]{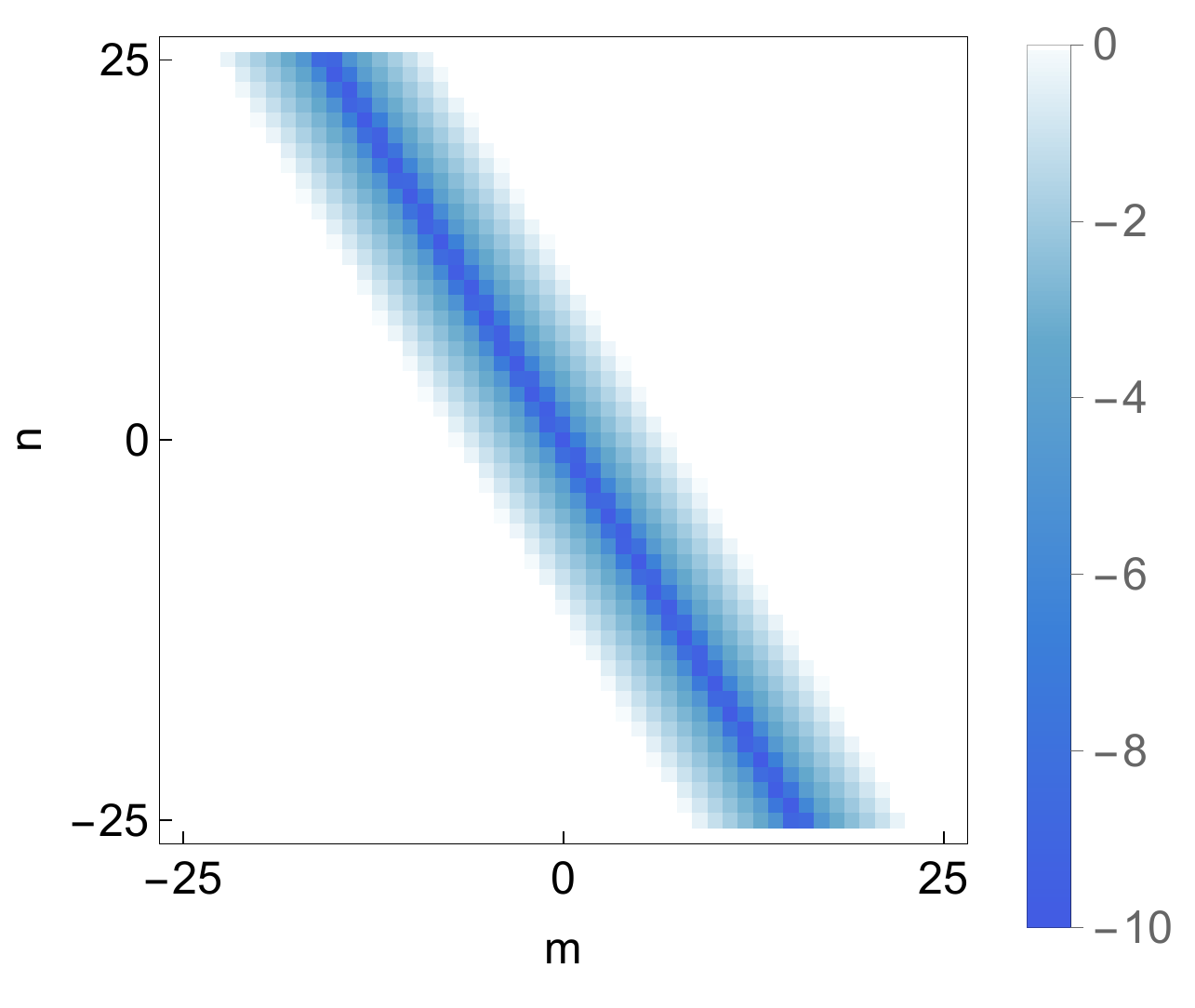}
\caption{(Color online) Plot of on-site potential in the case of $\bm{a} = (\frac{\sqrt{5}+1}{2}, 1)$ and $V(x)=-V e^{-x^2/w^2}$, where $V=10$, $w=1$, and $\delta=0$. Horizontal and vertical axes correspond to the $m$- and $n$-directions respectively.}
\label{onsite}
\end{center}
\end{figure}

Stimulated by the above-mentioned previous studies, it is natural to ask what happens if the additional diagonal lattice in the optical Lieb lattice is angled. Here, as a paradigmatic example, we consider a Fermi gas loaded in a 2D optical square lattice with a superlattice structure that is introduced obliquely. The lattice constant is taken to be unity. We study the topological properties of the system, which is described by the following Hamiltonian:

\begin{eqnarray}
\label{tbhamil}
\hat{\cal{H}}=&&\sum_{m,n} \left[ -t \left( \hat{c}_{m+1,n}^{\dag} \hat{c}_{m,n} + \hat{c}_{m,n+1}^{\dag} \hat{c}_{m,n}\right) + \mathrm{H.c.} \right] \nonumber \\
&&+ \sum_{m,n} \left[ V_{m,n} \hat{c}_{m,n}^{\dag} \hat{c}_{m,n} \right], \\
V_{m,n} &&= V(d), \\
d &&= \frac{\bm{a}}{|\bm{a}|} \cdot \left( \begin{array}{c} m-\delta \\ n\\ \end{array} \right),
\end{eqnarray}
where $\hat{c}_{m,n}^{\dag} (\hat{c}_{m,n})$ is the fermion creation (annihilation) operator, $t$ is the hopping strength, $\delta$ is the parameter of the real-space potential shift, $\bm{a} = (a_x, a_y)$ is the normal vector of the superlattice direction, and $d$ is the distance from the line expressed by $a_x (x-\delta) + a_y y = 0$.
The boundary conditions are taken as either open or periodic. Below, we take $t=1$ as the unit of energy. 

\subsection{Obliquely Introduced Single Well}

\begin{figure}[t]
\begin{center}
\includegraphics[width=9cm]{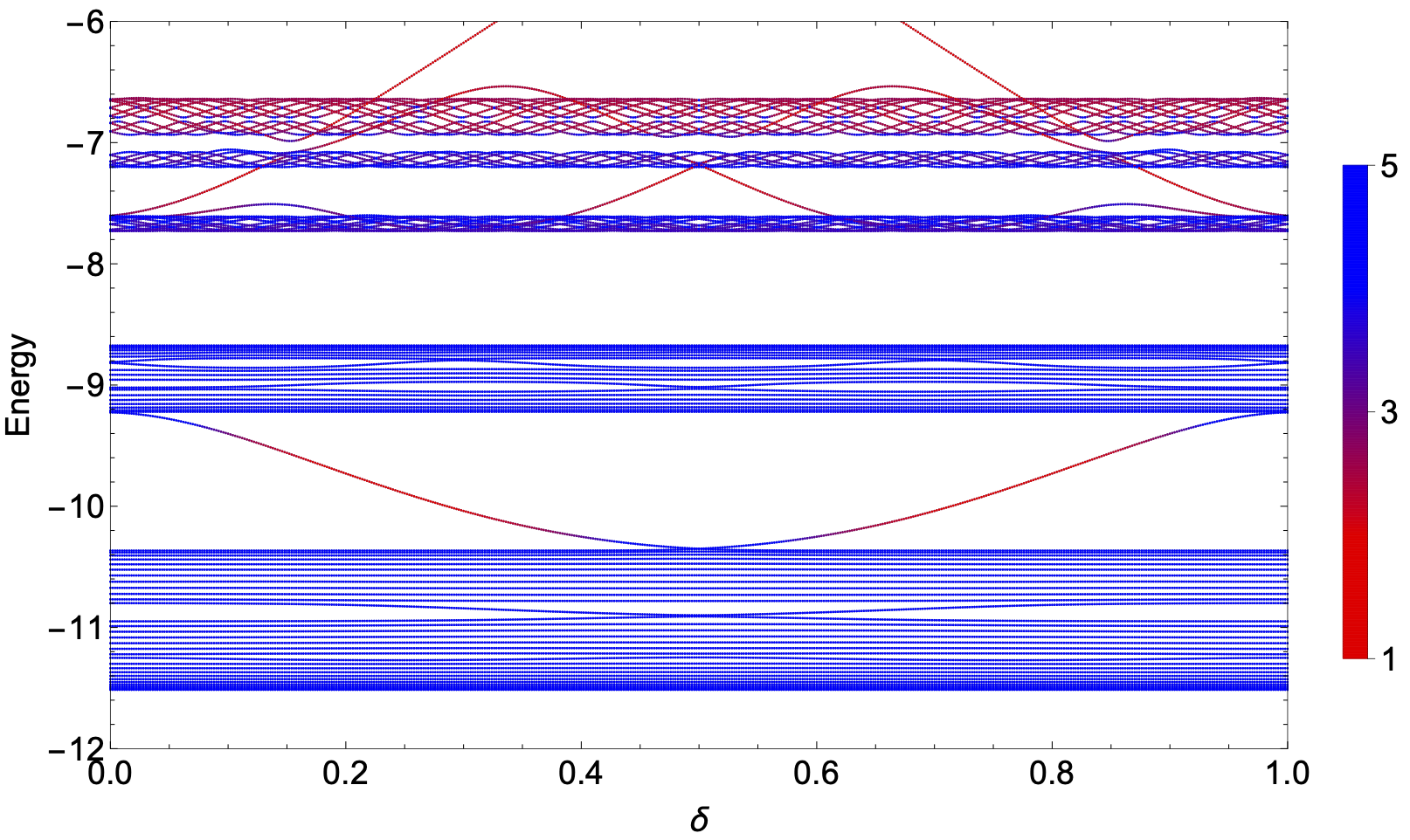}
\caption{(Color online) Energy spectrum plotted against the parameter $\delta$, which represents to the extent to which the obliquely introduced potential is shifted spatially. The on-site potential is given in Fig. \ref{onsite} with open boundary conditions. The bulk of the spectrum remains unchanged, with a few of the states crossing the band gap. The color shows the PR, which measures the spread of the eigenfunction.}
\label{spectrum}
\end{center}
\end{figure}
\begin{figure}
\begin{tabular}{cc}
\begin{minipage}{0.45\hsize}
\centering (a) \\
\flushright \includegraphics[width=4cm]{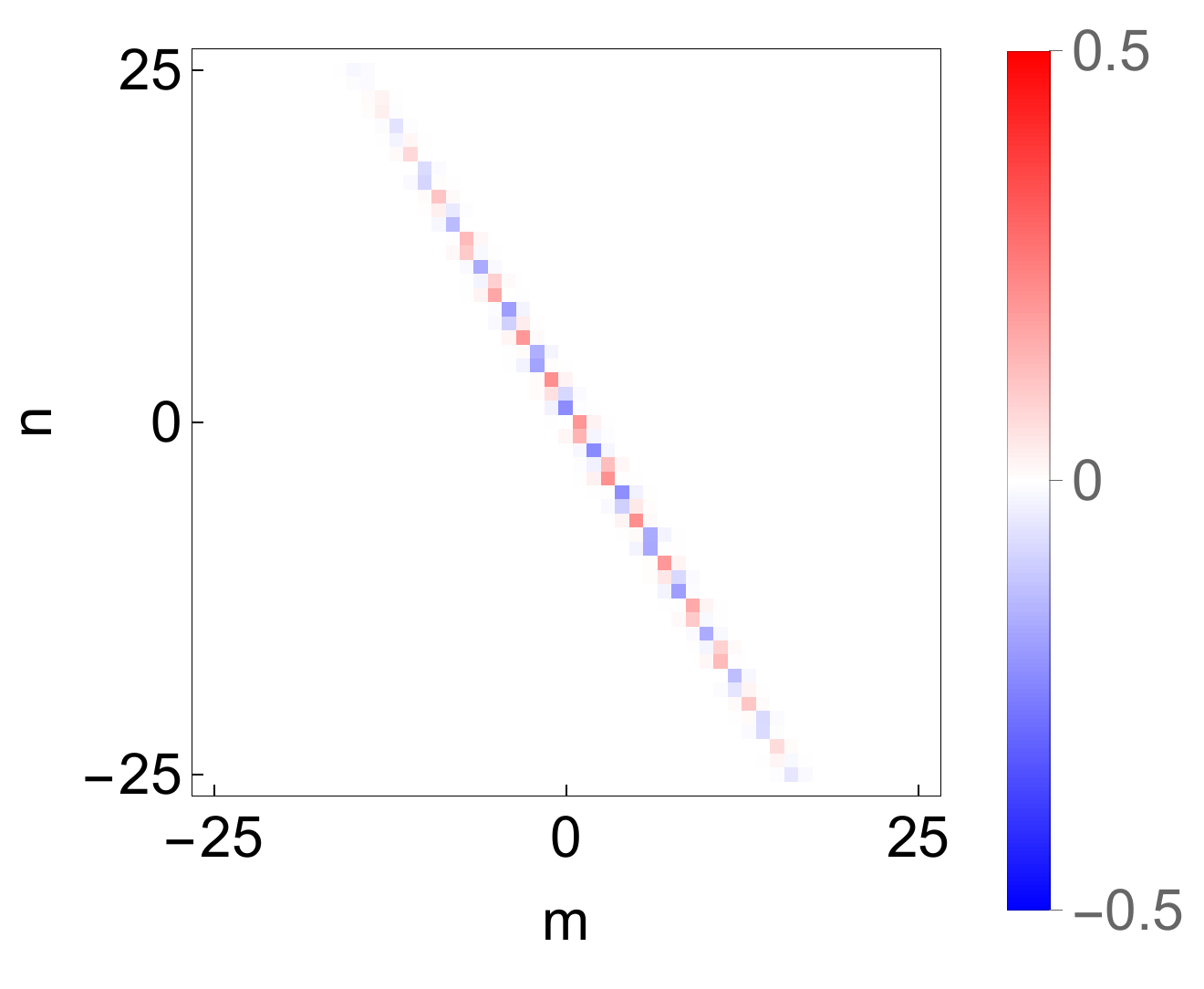}
 \end{minipage}
 \begin{minipage}{0.45\hsize}
 \flushleft (b) \\
\flushleft \includegraphics[width=4cm]{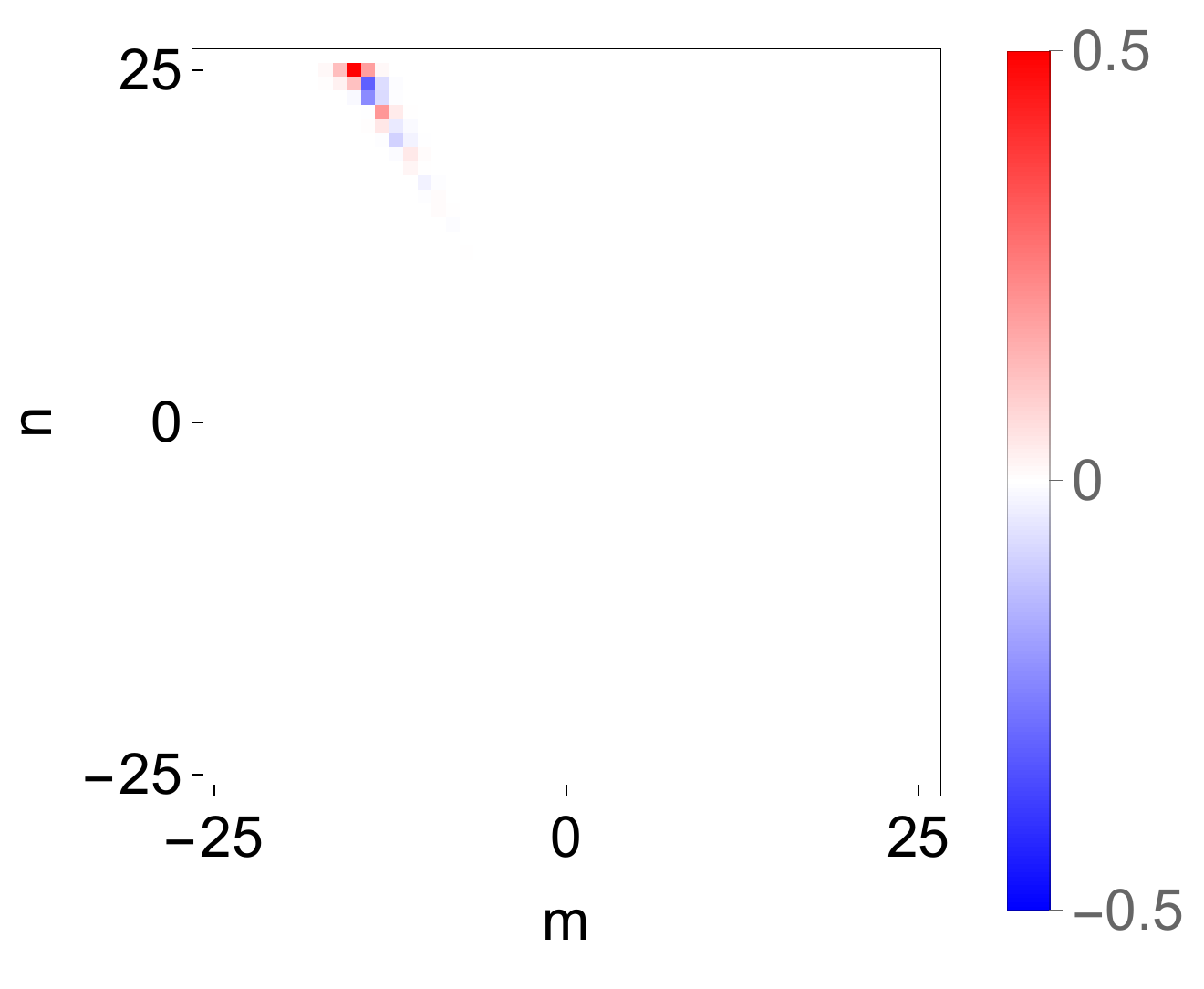}
 \end{minipage}
\end{tabular}
\caption{(Color online) Wave functions of the energy eigenstate in the $xy$ plane: (a) 30th and (b) 31st excited states in real space with $\delta = 0.75$. Whereas the wave functions are spread over the oblique potential in (a), they are localized at the edge in (b). Horizontal and vertical axes correspond to the $m$- and $n$-directions, respectively.}
\label{edgelocalized}
\end{figure}

First, we consider the case that $V(d)$ has a single dip, where the Hamiltonian has an oblique linear single well.  We take $V(d)=-V e^{-d^2/w^2}$, $\bm{a} = (\frac{\sqrt{5}+1}{2}, 1)$ as the linear-shaped potential, which restricts particles from moving around the potential in the low-energy region. The on-site potential in this system is plotted in Fig. \ref{onsite}. As a consequence of the overlap of the square lattice and the obliquely introduced potential, this model can be effectively considered as a 1D tight-binding model with a superlattice. This effective superlattice opens the energy gap. By analogy to the previous studies of topological aspects of a 1D superlattice\cite{PhysRevLett.108.220401,Kraus2012,JPSJ.83.083707}, the edge states will appear when we drive the phase of the superlattice, which corresponds to the parameter $\delta$ in this model. In Fig. \ref{spectrum}, we diagonalize the tight-binding model, which is described in Eq. (\ref{tbhamil}), with the system size $50 \times 50$ and under open boundary conditions. Figure \ref{spectrum} shows the plots for the bottom part of the energy spectrum since the upper part of the spectrum has no band gap because of the 2D nature of the wave function. We can see the existence of band-crossing states in the energy spectrum with open boundary conditions as a function of $\delta$ in Fig. \ref{spectrum}. 
The color in the figure shows the participation ratio (PR), which is defined as
\begin{equation}
I_{\psi} = \left( \sum_{m,n} |\psi_{m,n}|^4 \right)^{-1}.
\label{PR}
\end{equation}
A small PR indicates that the eigenfunction is well localized. 
Note that the bulk band is fixed because we choose an irrational number for tangent $\alpha$. Figures \ref{edgelocalized}(a) and 3(b) respectively show the wave functions of an in-band state and a band-crossing state. We confirm that the wave function of the band-crossing state is localized around the edge.

\begin{figure}
\begin{center}
\includegraphics[width=9cm]{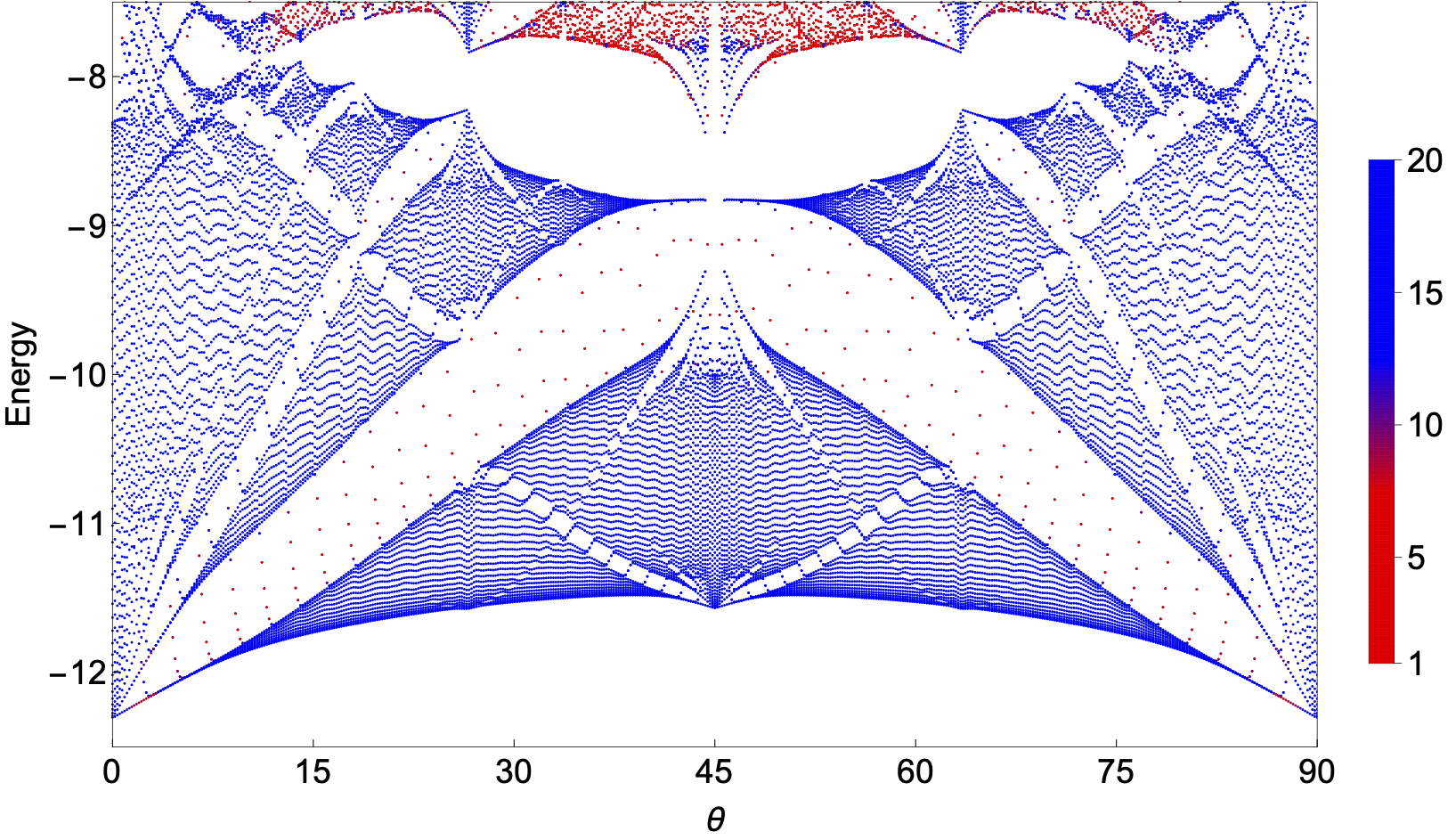}
\caption{(Color online) Butterfly structure in plots of the energy spectrum against the tilting angle $\theta$  in the case of $\bm{a} = (\cos \theta, \sin \theta)$, $V(x)=-V e^{-x^2/w^2}$, $V=10$, $w=1$, $\delta=0$. The color shows the PR, which is defined in Eq. (\ref{PR}).}
\label{butterfly}
\end{center}
\end{figure}

With changing $\bm{a}$, the distribution of the energy band also changes. Figure \ref{butterfly} shows the energy spectrum plotted against tilting angle $\theta$, which is related to $\bm{a}$ through the formula $\bm{a} \propto (\cos \theta, \sin \theta)$. We diagonalize the tight-binding model with the system size $50\times50$ and under open boundary conditions. For the same reason as in Fig. \ref{spectrum}, Fig. \ref{butterfly} shows the plots only for the bottom part of the energy spectrum. As shown in the figure, a butterfly-like structure appears in the energy spectrum plot. This structure is similar to the ``3D butterfly'' introduced in Refs. \onlinecite{Koshino2001} and \onlinecite{Koshino2002}. We will explain the origin of this similarity later. The color in Fig. \ref{butterfly} represents the PR in each eigenfunction, which is defined in Eq. (\ref{PR}). We can see that the eigenstates in the band gap are well localized, corresponding to edge states in topological systems.

The upper part of Fig. 4 also shows many localized states, but this is not the point we want to focus on because the wave functions in this region are no longer quasi-1D.

\subsection{Obliquely Introduced Superlattice}
As we have seen in the previous section, the band-crossing states and the edge-localized states appear when $V(d)$ is a single well, exemplifying the topological nature of the effective 1D system. We will now discuss topological properties in two dimensions when $V(d)$ is a periodic function. As a typical periodic function, we take $V(d) = V \cos (\lambda d)$. In this case, when we vary $\delta$ slowly, the Hamiltonian will be periodic in time, and this setup corresponds to the adiabatic transport, first proposed by Thouless\cite{Review1983}.

It is known that a mapping between a 1D superlattice system and a 2D Hofstadter model exists\cite{Kraus2012a}. Analogously, we can clearly understand the topological aspects of our model by mapping it to a 3D Hamiltonian. For simplicity, we start our discussion with the case that $a_x/a_y$ is a rational number. 
The Hamiltonian can be rewritten as follows:
\begin{eqnarray}
&&{\cal H}(\phi) \nonumber \\
&&=-t\sum_{m,n} \left[ \hat{c}^{\dag}_{m+1,n}(\phi)\hat{c}_{m,n}(\phi)+\hat{c}^{\dag}_{m,n+1} (\phi )\hat{c}_{m,n}( \phi) +\mathrm{H.c.} \right] \nonumber \\
 &&+ V \sum_{m,n} \cos \left( 2\pi \left( \frac{p_x}{q_x}m+\frac{p_y}{q_y}n \right)-\phi \right)\hat{c}^{\dag}_{m,n}(\phi)\hat{c}_{m,n}( \phi ),
\label{hamiltonian2D}
\end{eqnarray}
where $q_x$ and $q_y$ determine the periods for the $x$- and $y$-directions of the Hamiltonian, respectively. Also, we use $\phi \equiv 2\pi \delta p_x/q_x$ as a parameter controlling the phase of the cosine-type potential.
We can map this Hamiltonian to a 3D Hamiltonian by using the Fourier transformation,
\begin{eqnarray}
\hat{c}_{m,n}(\phi)=\cfrac{1}{\sqrt{2\pi L_z}}\sum_l e^{il\phi} \hat{c}_{m,n,l} \\
\hat{c}_{m,n,l}=\cfrac{1}{\sqrt{2\pi L_z}}\sum_l e^{-il\phi} \hat{c}_{m,n}(\phi).
\end{eqnarray}
Hamiltonian (\ref{hamiltonian2D}) will become
\begin{eqnarray}
{\cal H}_{\mathrm{3D}}&&=- t\sum_{m,n,l} \left[ \hat{c}^{\dag}_{m+1,n,l}\hat{c}_{m,n,l}+\hat{c}^{\dag}_{m,n+1,l}\hat{c}_{m,n,l} +\mathrm{H.c.} \right] \nonumber \\
&&+ \cfrac{V}{2} \sum_{m,n,l} \left[ e^{ 2\pi i \left( \frac{p_x}{q_x}m+\frac{p_y}{q_y}n \right)}\hat{c}^{\dag}_{m,n,l+1}\hat{c}_{m,n,l}+\mathrm{H.c.} \right] .
\label{hamiltonian3D}
\end{eqnarray}
This Hamiltonian describes a 3D tight-binding model on the cubic lattice with a homogeneous magnetic field whose direction is perpendicular to the $z$-axis but oblique in the $xy$-plane. This model has been previously studied, and it is known that the Hofstadter butterfly and the integer quantum Hall effect (IQHE) also appear in the 3D lattice\cite{PhysRevB.41.11417,PhysRevB.45.13488,Koshino2001, Koshino2002}. According to Refs. \onlinecite{Koshino2001} and \onlinecite{Koshino2002}, each gap in the 3D Hofstadter butterfly is characterized by two Chern indices. In the 3D model, these indices are proportional to the Hall conductivity. When we go back to our 2D model, these indices determine the topological pumping, just like Thouless pumping, and the bulk-edge correspondence shows that the amount of pumping and the number of edge states match each other. In this model, the amount of pumping in the $x$- and $y$-directions corresponds to the number of localized states at edges of $x$- and $y$-directions, respectively. Also, by comparing (\ref{hamiltonian2D}) and (\ref{hamiltonian3D}), we can see that potential strength $V$ corresponds to the hopping strength in the $z$-direction. These correspondences explain why the 3D butterfly appears in Fig. \ref{butterfly}.

Here, we have only considered a cosine-type potential, but a similar mapping is also possible for other types of potentials for which there are higher frequency components, and these components will be mapped to longer distance hoppings in the $z$-direction.  

\subsection{Diophantine Equation}
By mapping the Hamiltonian, we have seen that the Chern number characterizes the system, but how does the Chern number appear in observables? This question can be solved by considering the polarization. Since the derivative of the polarization is proportional to the current, the amount of pumping is proportional to the total amount of change in the polarization. Now, let us briefly summarize the theory of polarization.

According to Ref. \onlinecite{Resta2007}, the polarization has an integer ambiguity by a multiple of the lattice constant, but the change of the polarization in the course of a continuous change in the Hamiltonian can be defined without ambiguity.
When we write the wave vector of the electronic wave function as $\mathbf{k} = (k_x , k_y)$, the change in the polarization can be written as
\begin{equation}
\Delta \mathbf{P}_{\mathbf{i} \rightarrow \mathbf{f}}=\int_{\mathbf{i}}^{\mathbf{f}}\left(\partial_{\phi} \mathbf{P}\right) d \phi,
\label{Pif}
\end{equation}
\begin{equation}
\partial_{\phi} \mathbf{P}=\frac{e}{S} \sum_{n}^{\mathrm{occ}} \int_{\mathrm{BZ}} 2 \operatorname{Im}\left\langle\partial_{\phi} \psi_{n \mathbf{k}} | \partial_{\mathbf{k}} \psi_{n \mathbf{k}}\right\rangle d^{2} k,
\label{pol}
\end{equation}
where $\phi$ is the parameter of the Hamiltonian, $S$ is the area of the Brillouin zone, and the charge of an electron is $-e$. The expression for the integrand has the same form as the Berry curvature\cite{doi:10.1098/rspa.1984.0023}.

Following the above formula for the change in the polarization, we can define the amount of pumping even in infinite systems.
Namely an analogy with Eqs. (\ref{Pif}) and (\ref{pol}) enables us to express the amount of pumping in the $x$- and $y$-directions for our 2D system when $\phi$ is increased from $\phi_\mathrm{i}=0$ to $\phi_\mathrm{f} = \phi$ as
\begin{widetext}
\begin{equation}
P_x(\phi) = -\int_{0}^{\phi} d \phi' \frac{a_y}{2\pi}\int_{-2\pi/a_y}^{2\pi/a_y} d k_y \frac{a_x}{2\pi}  \int_{-2\pi/a_x}^{2\pi/a_x} d k_x 2 \operatorname{Im}\left\langle\partial_{\phi'} \psi_{n k_x} | \partial_{k_x} \psi_{n k_x}\right\rangle ,
\label{pol_chern_x1}
\end{equation}
\begin{equation}
P_y(\phi) = -\int_{0}^{\phi} d \phi' \frac{a_y}{2\pi}\int_{-2\pi/a_y}^{2\pi/a_y} d k_y \frac{a_x}{2\pi}  \int_{-2\pi/a_x}^{2\pi/a_x} d k_x 2 \operatorname{Im}\left\langle\partial_{\phi'} \psi_{n k_y} | \partial_{k_y} \psi_{n k_y}\right\rangle .
\label{pol_chern_y1}
\end{equation}
\end{widetext}
In the presence of the superlattice potential, the lattice periodicity is multiplied by $q_x$ and $q_y$ for the $x$- and $y$-directions, respectively. Therefore, the size of the first Brillouin zone shrinks by $(1/q_x) \times (1/q_y)$. This can be regarded as an analog of the magnetic Brillouin zone\cite{inverse}. We note that the signs in Eqs. (\ref{pol_chern_x1}) and (\ref{pol_chern_y1}) are opposite to the counterpart in Eq. (\ref{pol}) because of the sign of an electron charge.

Considering the case where $\phi$ is periodic with period $2\pi$, the amounts of pumping in $x$- and $y$-directions in one period for our 2D system are written as
\begin{widetext}
\begin{equation}
P_x(2\pi) = -\frac{a_y}{2\pi}\int_{-2\pi/a_y}^{2\pi/a_y} d k_y \left[ \frac{a_x}{2\pi} \int_{0}^{2\pi} d \phi \int_{-2\pi/a_x}^{2\pi/a_x} d k_x 2 \operatorname{Im}\left\langle\partial_{\phi} \psi_{n k_x} | \partial_{k_x} \psi_{n k_x}\right\rangle \right],
\label{pol_chern_x2}
\end{equation}
\begin{equation}
P_y(2\pi) = -\frac{a_x}{2\pi}\int_{-2\pi/a_x}^{2\pi/a_x} d k_x \left[ \frac{a_y}{2\pi} \int_{0}^{2\pi} d \phi \int_{-2\pi/a_y}^{2\pi/a_y} d k_y 2 \operatorname{Im}\left\langle\partial_{\phi} \psi_{n k_y} | \partial_{k_y} \psi_{n k_y}\right\rangle \right] .
\label{pol_chern_y2}
\end{equation}
\end{widetext}
The formula in the square brackets in Eqs. (\ref{pol_chern_x2}) and (\ref{pol_chern_y2}) has the same form as the definition of the Chern number on the torus created by $(\phi, k_x)$ and $(\phi, k_y)$, respectively. The Chern number is guaranteed to be an integer, but $P_x(P_y)$ is not necessarily an integer because it is given by the average of Chern numbers. However, if changing $k_y(k_x)$ does not close the band gap, then neither does the Chern number change, so $P_x(P_y)$ is also an integer. To summarize, in an adiabatic time-periodic system, the amount pumped in the $x$-direction during one period is represented by the integral of the change in polarization in the $x$-direction, which corresponds to the average in the $k_y$-direction of the Chern numbers for $(\phi, k_x)$. 

\begin{table}[t]
\begin{tabular}{c|cccccccccc}
$q_x$ & 2 & 3 & 2 & 3 & 4 & 2 & 3 & 4 & 5 & 6 \\ 
$q_y$ & 3 & 4 & 5 & 5 & 5 & 7 & 7 & 7 & 7 & 7 \\ \hline
$u$ & 1 & 1 & 1 & -1 & 1 & 1 & 1 & -1 & -2 & 1 \\
$v$ & -1 & -1 & -2 & 2 & -1 & -3 & -2 & 2 & 3 & -1
\end{tabular}
\caption{Several examples of the solution of the Diophantine equation (\ref{3Ddiop1}) with the smallest absolute value in the case of $p_x=p_y=1$, $r=1$. }
\label{cherntable}
\end{table}

As we have mentioned above, in the tight-binding approximation, we can map our model to a 3D cubic lattice model with an obliquely introduced homogeneous magnetic field in Eq. (\ref{hamiltonian3D}), and the amount of Thouless pumping in our model corresponds to the quantized Hall conductance in the cubic lattice model. In Eq. (\ref{hamiltonian3D}), $x$ and $y$ are cyclic coordinates, so that the wave function can be written as $\Psi_{lmn}=e^{ik_xm+ik_yn}F_l$, where $k_x$ and $k_y$ are the Bloch wave numbers along the $x$- and $y$-directions, respectively. Then the Schr\"{o}dinger equation will be
\begin{equation}
t(F_{l-1}+F_{l+1})-V\{ \cos (2\pi lp_x/q_x+k_x)+\cos (2\pi lp_y/q_y+k_y) \} F_l = EF_l.
\end{equation}
This equation is equivalent to a 3D version of the Harper equation. By analyzing this equation, the following Diophantine equation can be obtained \cite{PhysRevB.45.13488}:
\begin{equation}
\frac{r}{Q} = s + u\frac{p_x}{q_x} + v\frac{p_y}{q_y} .
\label{3Ddiophantos}
\end{equation}
Here, $r$ is the number of filled bands and $Q$ is the least common multiple of $q_x$ and $q_y$. $u$ and $v$ correspond to the amounts of pumping when $\varphi$ changes from $0$ to $2\pi/Q$. 
The detailed derivation of this equation is shown in Appendix \ref{diotight}.

\begin{figure}
\flushleft \hspace{4.5cm} (a) \\
\centering \includegraphics[width=6.5cm]{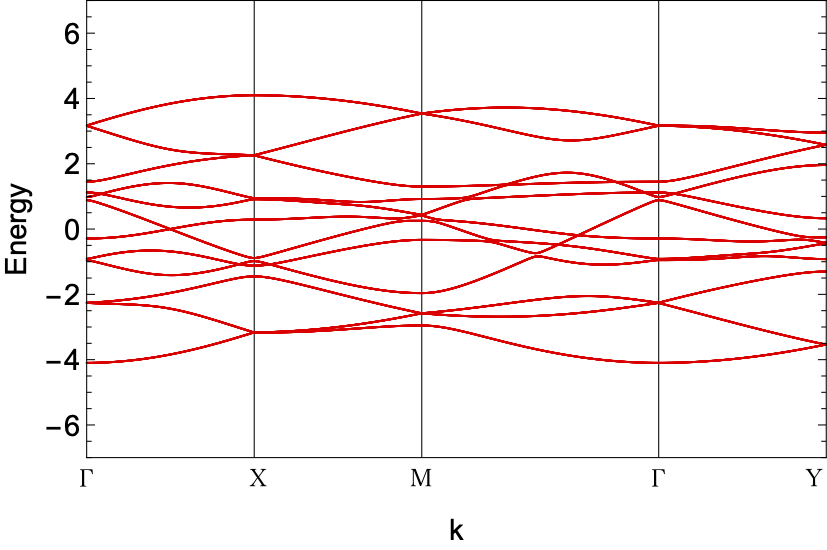} \\
\flushleft \hspace{4.5cm} (b) \\
\centering \includegraphics[width=6.5cm]{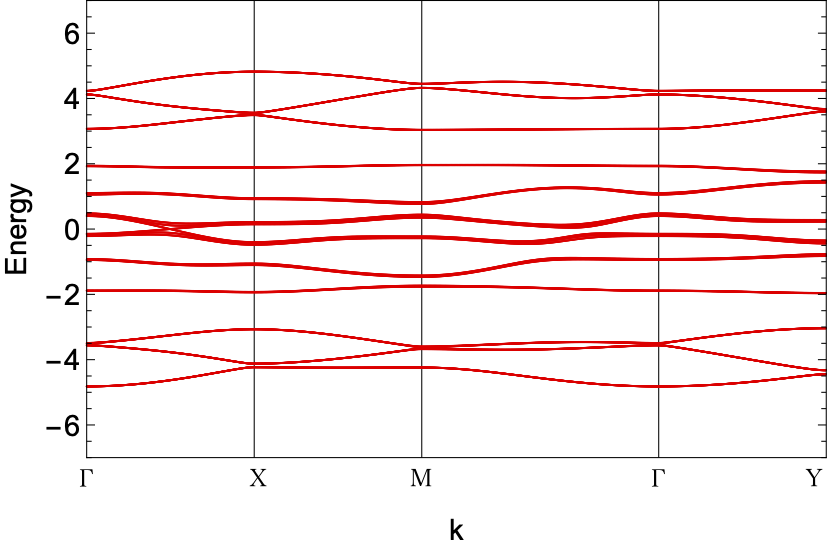} \\
\flushleft \hspace{4.5cm} (c) \\
\centering \includegraphics[width=6.5cm]{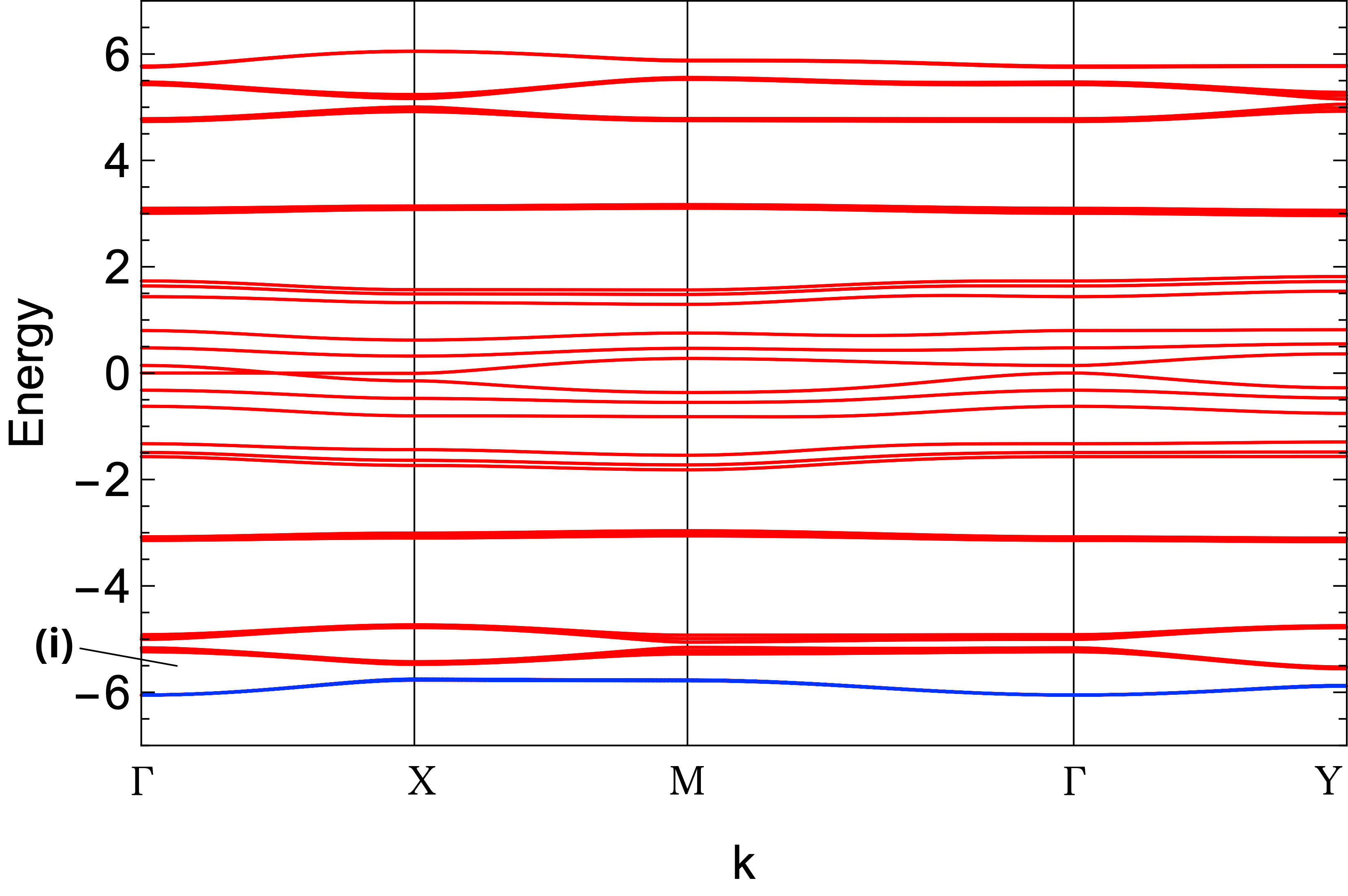}
\caption{(Color online) Band structures computed using the tight-binding approximation with $q_x = 3, q_y = 4$, and (a) $V=1.0$, (b) $V=3.0$, and (c) $V=5.0$. The blue band in (c) corresponds to the lowest band and the symbol (i) indicates the lowest band gap. The high-symmetry points in the first Brillouin zone are defined as $\Gamma=(0,0), X=(\pi,0), Y=(0,\pi)$, and $M=(\pi,\pi)$. }
\label{TBband}
\end{figure}

Below we focus on the case where $p_x = p_y = 1$ and $q_x, q_y$ are coprime. Then Eq. (\ref{3Ddiophantos}) can be transformed to 
\begin{equation}
r \equiv u q_y + v q_x \ (\bmod \ Q).
\label{3Ddiop1}
\end{equation}
Basically, we are concerned with the lowest band, so we should consider the case $r=1$. Table \ref{cherntable} shows the solution of Eq. (\ref{3Ddiop1}) in several cases.

\begin{figure}[t]
\centering \includegraphics[width=7cm]{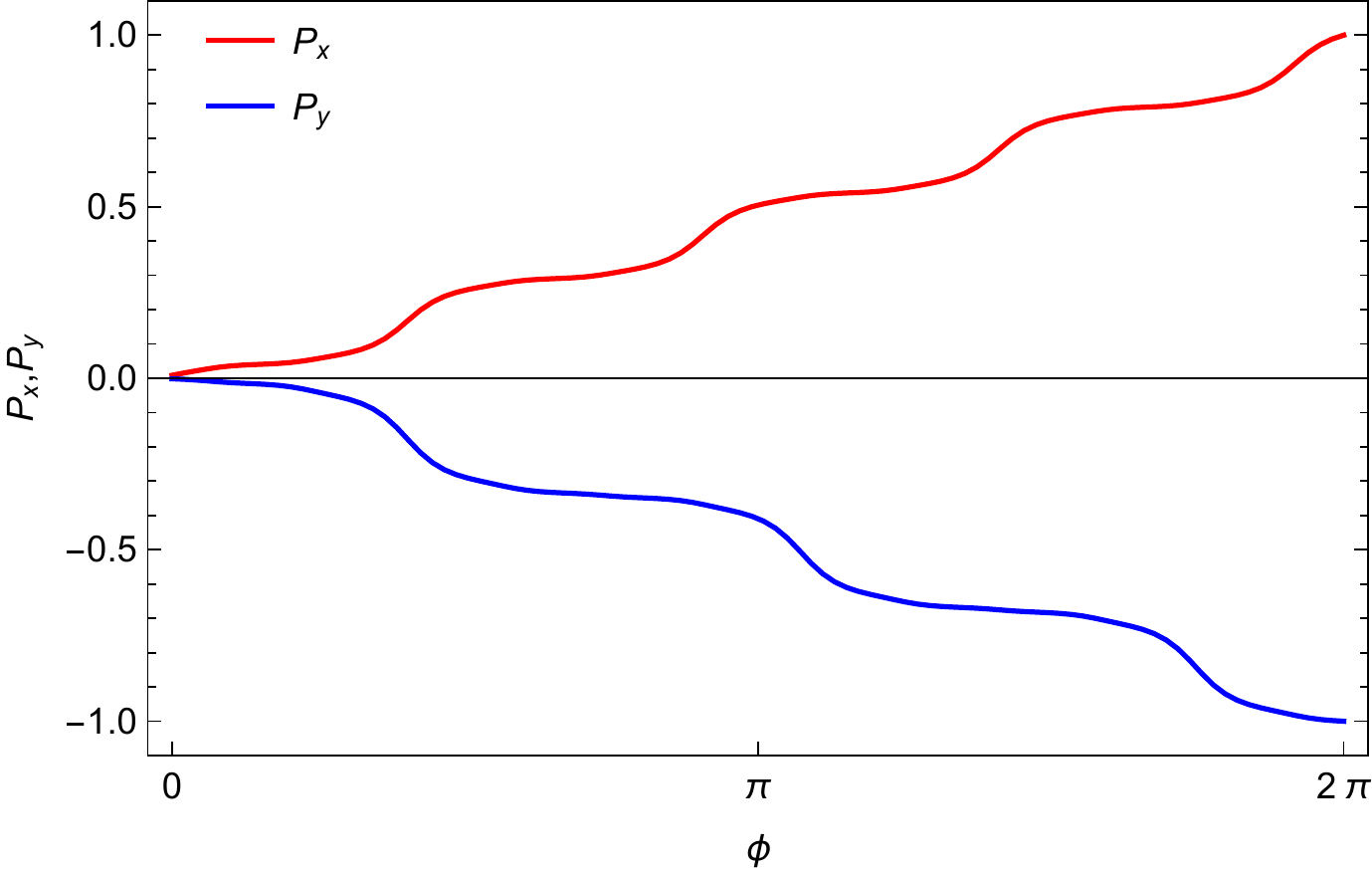}
\caption{(Color online) Amount of pumping, which is calculated by integrating the Berry curvature and using the tight-binding approximation with $q_x = 3, q_y = 4, V=5.0$.}
\label{3-4TB}
\end{figure}

\subsection{Band Structure and Pumped Amount}

In the previous section, we showed that the amount of pumping obeys a Diophantine equation.
Now we show a concrete example below. In particular, we focus on the case where $q_x=3, q_y=4$.

Figure \ref{TBband} shows how the band changes when the magnitude of the modulation $V$ of the onsite potential is changed. Figures \ref{TBband}(a)--(c) are the band structures for $V = 1.0, 3.0, 5.0$, respectively. 
When $V$ is large, the lowest band gap (i) is opened, which can be observed in Fig. \ref{TBband}(c). 
On the contrary, when $V$ is small, the Berry curvature cannot be calculated well since the band gap is small. The smaller the band gap, the larger the number of meshes required for the Fukui--Hatsugai--Suzuki method. 
The reason why the band gap is so small is that the higher order perturbations open the gap.
It is not likely that the band gap is closed in the process of increasing $V$, so it seems that the gap is open even if $V$ is very small, but is too small to be detected numerically.

Here we investigate the amount of pumping for the lowest band in Fig. \ref{TBband}(c). We apply the Fukui--Hatsugai--Suzuki method\cite{doi:10.1143/JPSJ.74.1674} to calculate the amount of pumping, which can be obtained by integrating the Berry curvature\cite{doi:10.1098/rspa.1984.0023}. The specific expression for the amount of pumping is given in Eqs. (\ref{pol_chern_x1}) and (\ref{pol_chern_y1}). 
Figure \ref{3-4TB} shows the amount of pumping for large $V$, which is calculated using Eqs. (\ref{pol_chern_x1}) and (\ref{pol_chern_y1}). The amount of pumping after one cycle, $P_x=1$ and $P_y=-1$, agrees with the solution of the Diophantine equation shown in Table \ref{cherntable}.

\section{Continuum Model of 2D Thouless Pumping}

\subsection{Model}

In light of the analogy of Thouless pumping and the IQHE, the results in the previous section suggest that the amount of transport, as the Hamiltonian parameter $\phi$ changes its value from 0 to $2\pi$, is quantized and can be expressed in terms of the Chern number. Since the discussion in the previous section is based on the tight-binding model, it is desirable to extend our discussion to continuum systems so as to apply the results to cold atom systems. To treat the system more precisely and to see the feasibility of the previous model in the experimental setups, we next consider a 2D continuum model.

To be specific, we consider the following Hamiltonian:
\begin{eqnarray}
\label{continuum1}
H_{\mathrm{cont}} &&= \int dx dy \psi^\dagger (x,y) \biggl[ -\cfrac{\hbar^2}{2m} \cfrac{\partial^2}{\partial x^2} -\cfrac{\hbar^2}{2m} \cfrac{\partial^2}{\partial y^2} \nonumber \\
&&+ V_x(x,y) + V_y(x,y) + V_{\mathrm{SL}}(x,y) \biggl] \psi(x,y), \\
&&V_j(x,y) = 2V_j \cos(q_j j) \ \ \ \ (j = x,y), \\
&&V_{\mathrm{SL}}(x,y) = 2V_{\mathrm{SL}} \cos(p_x x + p_y y - \varphi), 
\label{continuum2}
\end{eqnarray}
where $\psi(x,y)$ is the field operator of a particle, $\hbar$ is Planck's constant, and $m$ is the mass of a particle. There are three types of potentials in this system: $V_x$, $V_y$, and $V_{\mathrm{SL}}$. $V_x$ and $V_y$ form the 2D square lattice, and $V_{\mathrm{SL}}$ forms the superlattice oblique to the square lattice. Henceforth, we take $\hbar^2/m$ as the unit of energy.

\begin{figure}[t]
\flushleft \hspace{4.5cm} (a) \\
\centering \includegraphics[width=6.5cm]{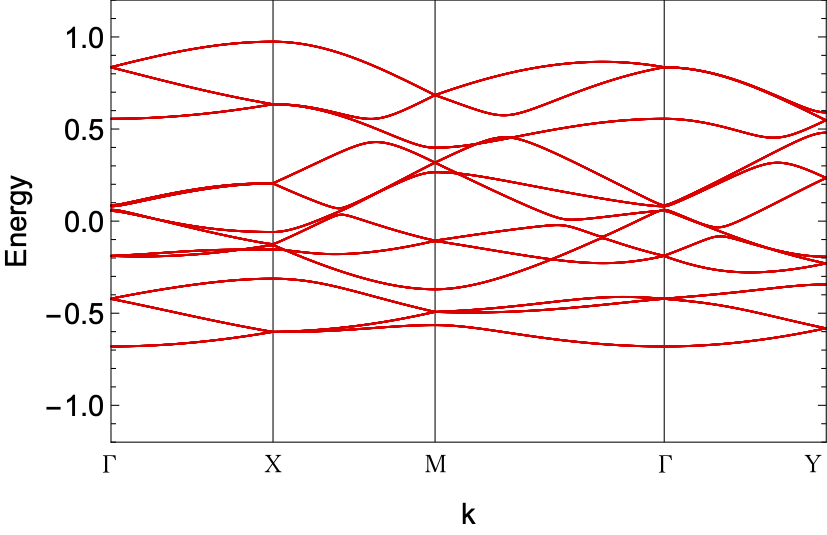} \\
\flushleft \hspace{4.5cm} (b) \\
\centering \includegraphics[width=6.5cm]{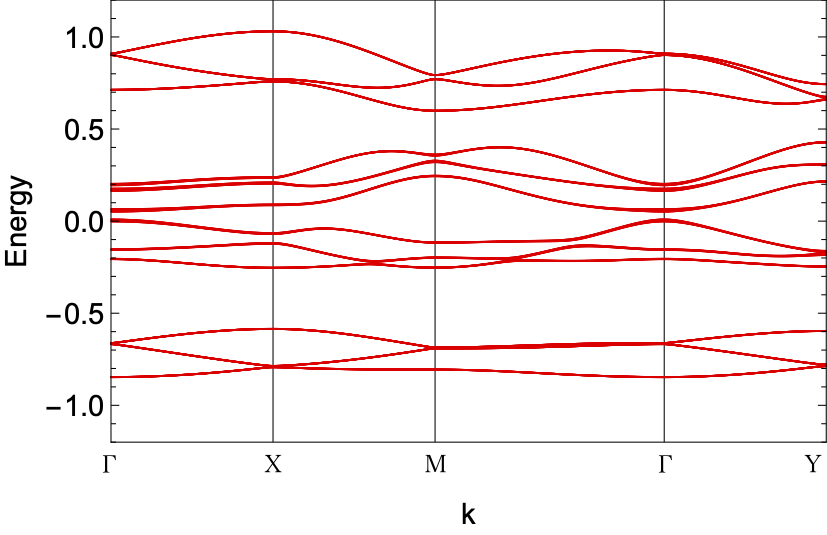} \\
\flushleft \hspace{4.5cm} (c) \\
\centering \includegraphics[width=6.5cm]{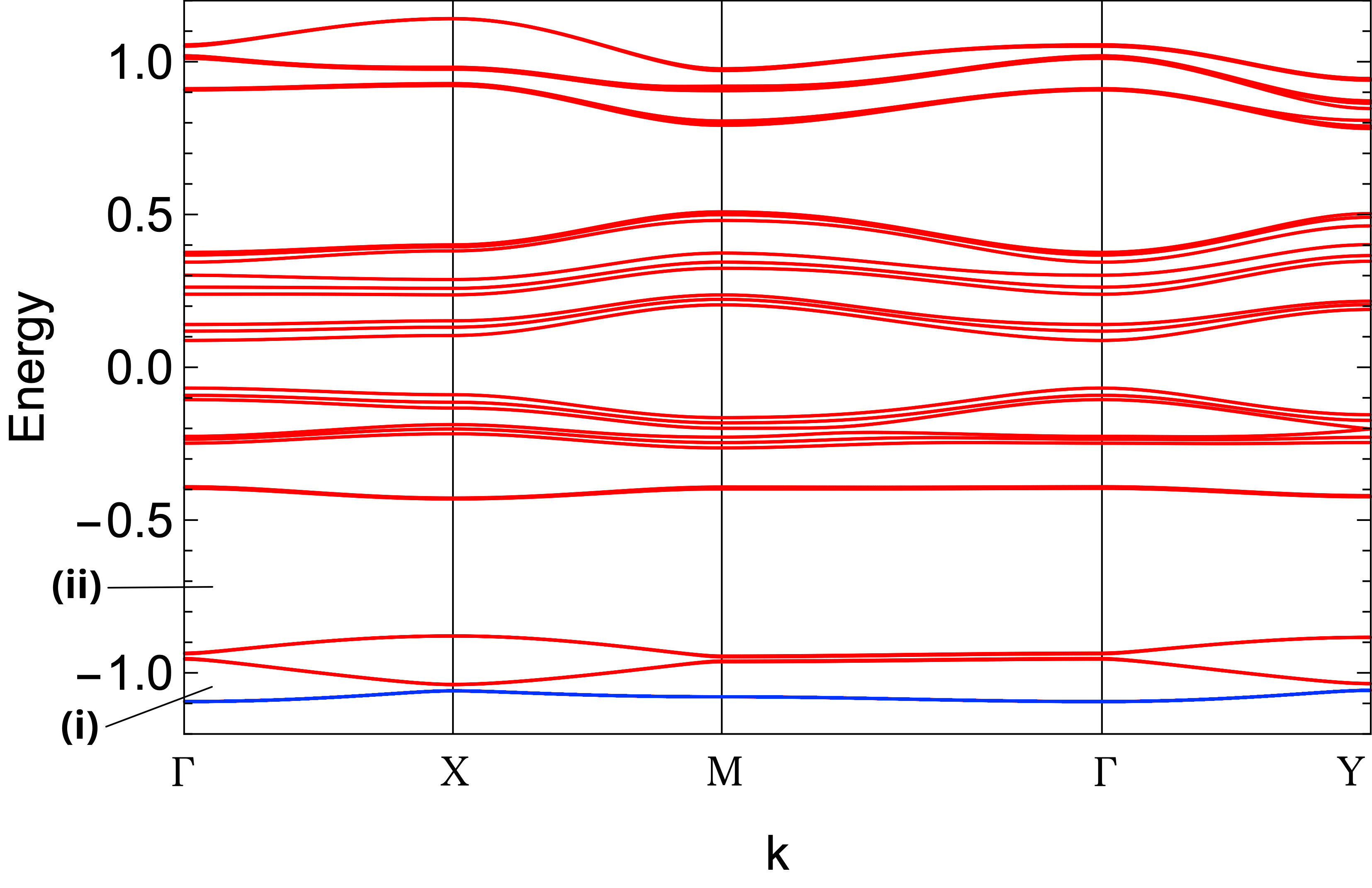}
\caption{(Color online) Band structure computed using the plane-wave approximation with $q_x = 3, q_y = 4, V_x=V_y=1.0$ and (a) $V_{\mathrm{SL}}=0.1$, (b) $V_{\mathrm{SL}}=0.3$, and (c) $V_{\mathrm{SL}}=0.5$. The blue band in (c) corresponds to the lowest band. The symbols (i) and (ii) respectively indicate the lowest band gap and the band gap between the second and third excited states. The high-symmetry points in the first Brillouin zone are defined as $\Gamma=(0,0), X=(\pi,0), Y=(0,\pi)$, and $M=(\pi,\pi)$.}
\label{PWband}
\end{figure}

\subsection{Plane-Wave Approximation}

\subsubsection{Band Structure}

We now apply the plane-wave approximation to the Hamiltonian introduced in Eqs. (\ref{continuum1})--(\ref{continuum2}) to calculate the eigenenergies and the wave functions directly. 
First, we apply the Fourier transformation
\begin{equation}
\psi_{k_x, k_y} = \cfrac{1}{\sqrt{L_x L_y}} \int dx dy e^{ik_x x+ik_y y}\psi(x,y)
\end{equation}
 to Eq. (\ref{continuum1}). Then, we obtain the following Hamiltonian:

\begin{equation}
\begin{split}
H &= \sum_{0\leq k_x,k_y < 1} \sum_{n_x,n_y = -\infty}^{\infty} \Bigl[ \\
&\cfrac{1}{2} \left((k_x+n_x)^2+(k_y+n_y)^2\right)\psi_{n_xn_y}^\dagger \psi_{n_xn_y} \\ 
&+\left( V_x \psi_{n_xn_y}^\dagger \psi_{n_x+q_x,n_y} + V_y \psi_{n_xn_y}^\dagger \psi_{n_x,n_y+q_y} \right. \\
&\left. + V_{\mathrm{SL}} e^{-i\varphi}\psi_{n_xn_y}^\dagger \psi_{n_x+p_x,n_y+p_y} + \mathrm{H.c.} \right) \Bigr].
\end{split}
\label{plane}
\end{equation}
When the condition $V_x, V_y, V_{\mathrm{SL}} \ll 1$ is satisfied, it is possible to diagonalize Eq. (\ref{plane}) numerically by introducing the cutoff for wavenumbers $k_x$ and $k_y$. 

Figure \ref{PWband} shows the band structure when the plane-wave approximation is used. In order to calculate the wave function under the plane-wave approximation, we introduce the cutoff of wavenumber $k$. To be specific, we only consider small $n_x$ and $n_y$ satisfying the condition $|n_x|, |n_y| \leqq 12$ in Eq. (\ref{plane}).
The band structure is similar to that in Fig. \ref{TBband} when $V_{\mathrm{SL}}$ is small, but the gap begins to increase as $V_{\mathrm{SL}}$ increases, and the first gap is clearly open in the plot for $V_\mathrm{SL} = 0.5$. Even in this model, it is unlikely that a phase transition accompanied by gap formation occurs somewhere, so we expect that a gap opens even if $V_{\mathrm{SL}}$ is small. When $V_{\mathrm{SL}}$ is small, the gap is formed by a higher order perturbation, which is probably not visible in the figure.

\subsubsection{Pumped Amount and Diophantine Equation}

\begin{figure}[t]
\vspace{-1.5cm}
\centering \includegraphics[width=7cm]{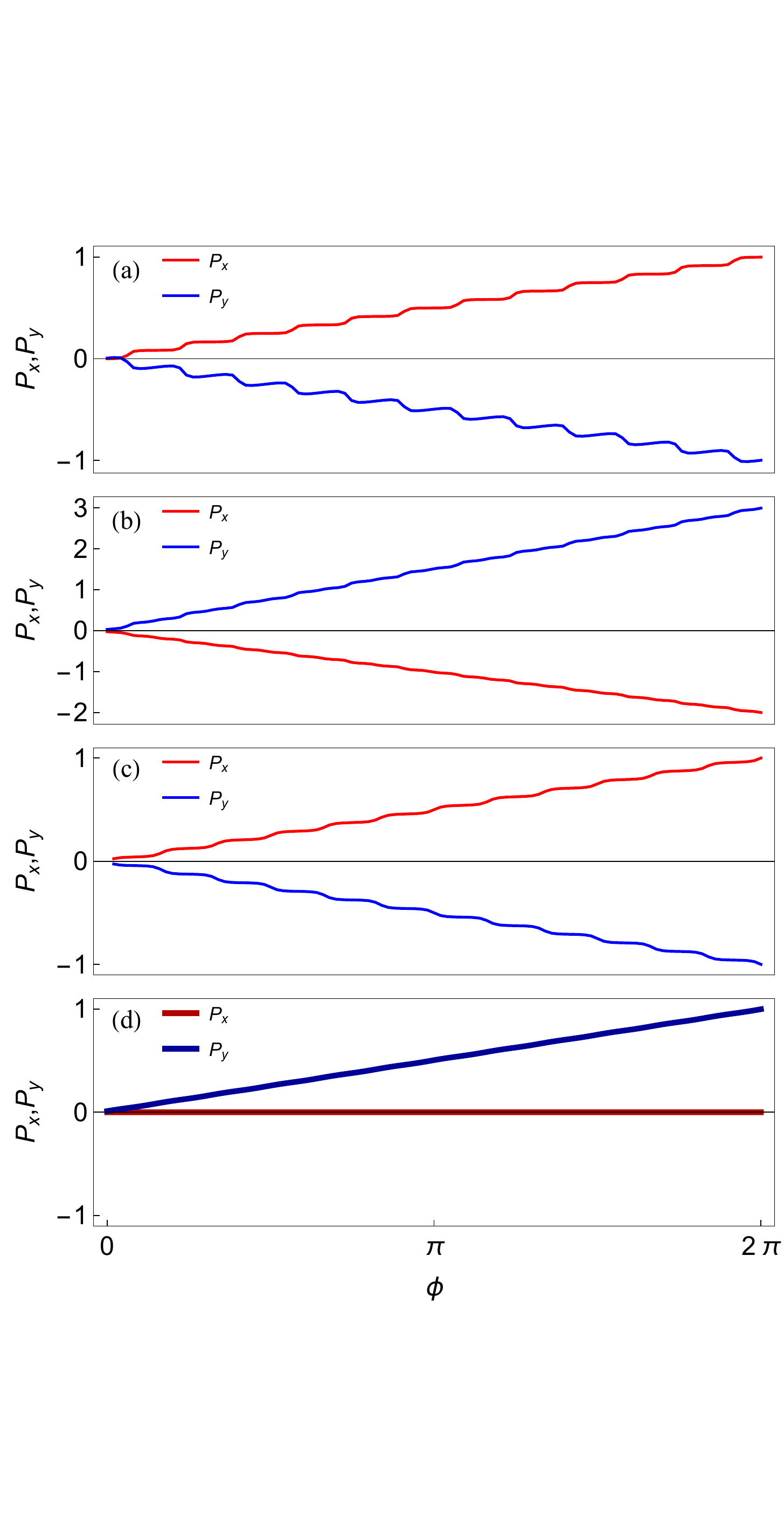} 
\vspace{-1.5cm}
\caption{(Color online) Amount of pumping for the ground state and the first and second excited states, calculated by integrating the Berry curvature and using the plane-wave approximation with $p_x = 3, q_x = 1, p_y = 4, q_y = 1, V_x = V_y = 2.0, V_{\mathrm{SL}}=0.5$. The bold line in (d) shows the total amount of pumping of these three states. Red and blue lines correspond to the $x$- and $y$-directions, respectively.}
\label{3-4PW}
\end{figure}

Figure \ref{3-4PW}(a) shows the amount of pumping for the ground state under the condition $q_x = 3, q_y = 4, V_x=0.5, V_y=0.5, V_{\mathrm{SL}}=0.25$. Since we use the plane-wave approximation to calculate wave functions, there are an infinite number of bands. Only the low-lying $q_x q_y$ bands are plotted. In this case, a single band corresponds to the $1/(q_x q_y)$ filling. The amount of pumping after one cycle is -1 in the $x$-direction and +1 in the $y$-direction, which is consistent with the solution of the Diophantine equation. Also, the pumping proceeds stepwise in 12 steps. This is consistent with $q_x q_y = 12$. 
In general, we suppose that the amount of pumping calculated in the plane-wave approximation follows the same Diophantine equation as in the tight-binding limit. The justification of the equation is given in Appendix B.

We here focus on the fact that the band gap between the second and third excited states is large in Fig. \ref{PWband}, requiring us to treat the low-lying states below the gap carefully. Figures \ref{3-4PW}(b) and \ref{3-4PW}(c) show the amount of pumping of the first excited state and the second excited state, respectively. Also, Fig. \ref{3-4PW}(d) shows the total amount of pumping of the states in Figs. \ref{3-4PW}(a)--\ref{3-4PW}(c). As mentioned earlier, the amount of pumping is also represented by the Diophantine equation (\ref{3Ddiophantos}) within the plane-wave approximation. Concretely, the amount of pumping when the corresponding state is located from the bottom to the $r$th state is the solution of Eq. (\ref{3Ddiophantos}). Therefore, if we denote the $n$th pair of Chern numbers from the bottom as $(C_x, C_y)$ and the solution of Eq. (\ref{3Ddiophantos}) for $r=n$ as $u_n, v_n$, we have $C_x = u_n-u_{n-1}$, $C_y = v_n-v_{n-1}$. Table \ref{cherntable2} shows the solutions of Eq. (\ref{3Ddiop1}) for $r=0,1,2,3$ of Eq. (\ref{3Ddiophantos}) and the difference between them for $q_x=3, q_y=4$. The amount of pumping after one period ($\phi=2\pi$) corresponds to the differences between the adjacent solutions of the Diophantine equation, $(1,-1)$, $(-2,3)$, and $(1,-1)$, which are listed in Table \ref{cherntable2}. These are consistent with Fig. \ref{3-4PW}. Also, Fig. \ref{3-4PW}(d) corresponds to the solution for $r=3$ in Table \ref{cherntable2}.

\begin{table}[t]
\begin{tabular}{c|cccccccccc}
$r$ & 0 &  & 1 &  & 2 &  & 3 \\ \hline
$u$ & 0 & $\xrightarrow{+1}$ & 1 &  $\xrightarrow{-2}$ & -1 &  $\xrightarrow{+1}$ & 0 \\ 
$v$ & 0 & $\xrightarrow{-1}$ & -1 &  $\xrightarrow{+3}$ & 2 &  $\xrightarrow{-1}$ & 1
\end{tabular}
\caption{Solutions of the Diophantine equation (\ref{3Ddiop1}) for different $r$ and the differences between them  in the case of $q_x=3, q_y=4$.}
\label{cherntable2}
\end{table}

We suppose that the wide band gap in $r=3$ is generated by a first-order perturbation, while other gaps are generated by higher order perturbations. The bold lines in Fig. \ref{3-4PW}(d) show the amount of pumping when the states are filled up to this wide band gap. In the present case, instead of dealing only with the lowest band, we sum up the contributions of the three bands from the bottom, as already mentioned. Then, the pumping in the $x$-direction becomes zero, and only the pumping in the $y$-direction remains. 

We also compute the Chern numbers for various cases with different $q_x$ and $q_y$. These include the case of $(q_x, q_y) = (5,7)$, where both $C_x$ and $C_y$ have absolute values of 2 or larger. In every case, $C_x$ and $C_y$ obey the Diophantine equation (\ref{3Ddiophantos}).
These results validate the claim that a pair of Chern numbers are determined by the Diophantine equation from the calculation of the Berry curvature.

\subsection{Effects of Harmonic Trap}

\begin{figure}[t]
\centering \includegraphics[width=7cm]{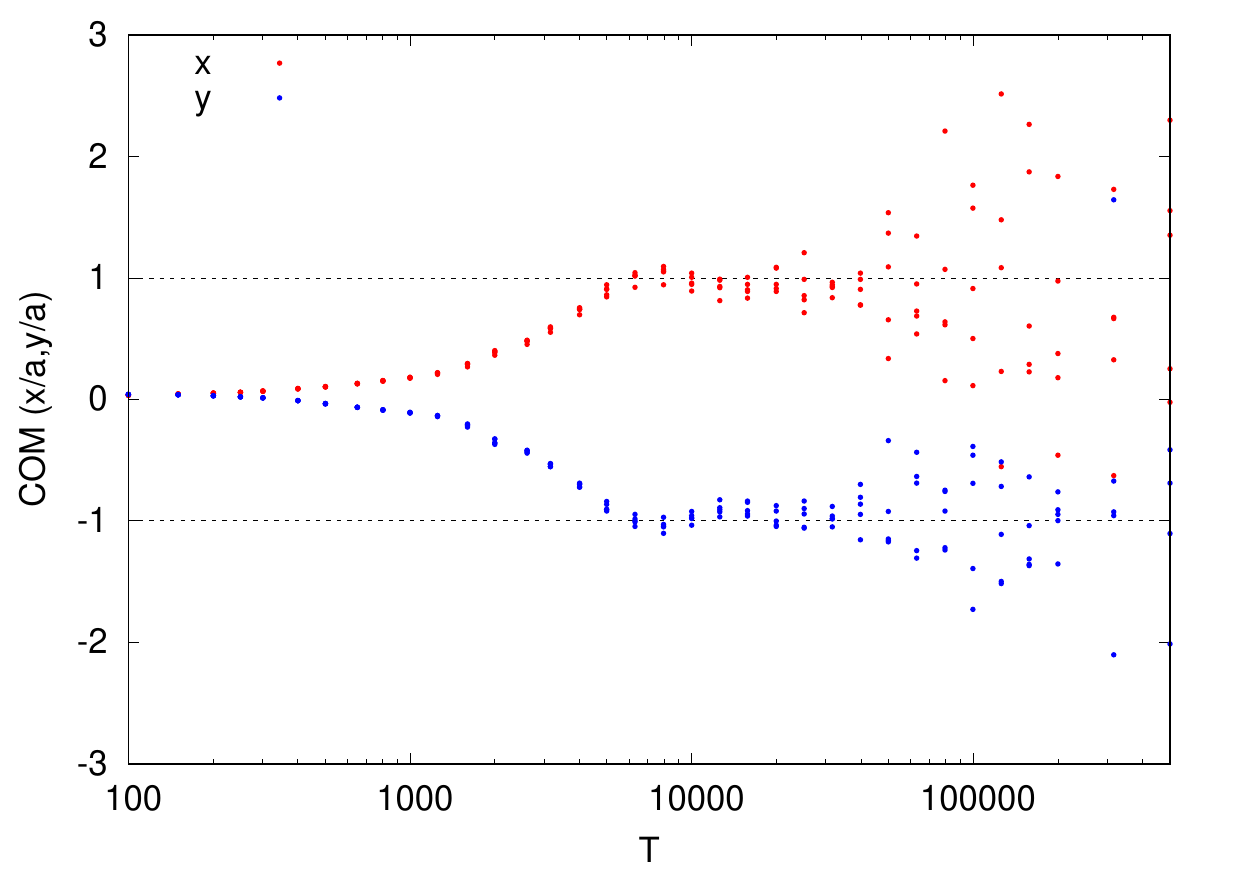}
\caption{(Color online) COM of each state plotted against the time period $T$ for the lowest five states in the case of $q_x = 3, q_y = 4$, $V_{\mathrm{L}}=1.5$, $V_{\mathrm{SL}}=0.5$, $V_{\mathrm{trap}}=0.00001$. }
\label{harmeff}
\end{figure}

So far, we have studied the continuum model (\ref{continuum1}) in the infinite system having $q_x \times q_y$-site periodicity with an integer number of filled bands. We have seen that the amount of pumping is quantized and the Chern numbers obey the Diophantine equation in the infinite system. In this section, we discuss the effect of a harmonic trap on the continuum model. 
We consider the following Hamiltonian, where the harmonic potential is added as a trap to the continuum model Hamiltonian (\ref{continuum1}), and simulate the time evolution of the lowest-energy states and their center of mass (COM).

\begin{eqnarray}
\label{timedepham}
H = H_{\mathrm{cont}}(x,y,t)+H_{\mathrm{trap}}(x,y) \\
H_{\mathrm{trap}} = V_{\mathrm{trap}}(x^2+y^2) \\
i\frac{\partial}{\partial t}\Psi (t) = H \Psi (t)
\end{eqnarray}

For the initial state, we employ the eigenstates calculated for the initial superlattice potential. We set $\phi= 0$ when one of the lattice minima is on the minimal line of the superlattice potential. Starting from $t = 0$, we solve the time-dependent Schr\"{o}dinger equation using the infinitesimal time evolution operator while the superlattice potential is changing. We set the mesh size to $\Delta x=0.2$ and the time step to $\Delta t = 0.1$.

When the harmonic potential exists, the difference in the COM of the wave functions of the occupied states corresponds to the amount of pumping. Let $(x_0, y_0)$ be the initial COM at $t=0$. These are not necessarily 0. We will plot the change in the COM, which is expressed as $\Delta x=x-x_0$, $\Delta y=y-y_0$. Below, we take the average COM of the five lowest energy states. Even if we change the number of states, the change in the COM does not significantly affect the amount of pumping owing to the harmonic potential, but in this case, we set the number of states to five to avoid edge effects. We can discuss the pumping behavior in terms of the computed COM.

In the presence of a harmonic trap, the pumping behavior turns out to be completely different between two regions. One is the weak superlattice region, in which $V_{\mathrm{SL}}$ is small compared with $V_{\mathrm{L}}=V_x=V_y$ (which is called the Hofstadter region), and the other is the strong superlattice region, in which $V_{\mathrm{SL}}$ is large (which is called the rectifying region). 

\subsubsection{Hofstadter Region}

\begin{figure}[t]
\begin{center}
(a) \\ \includegraphics[width=7cm]{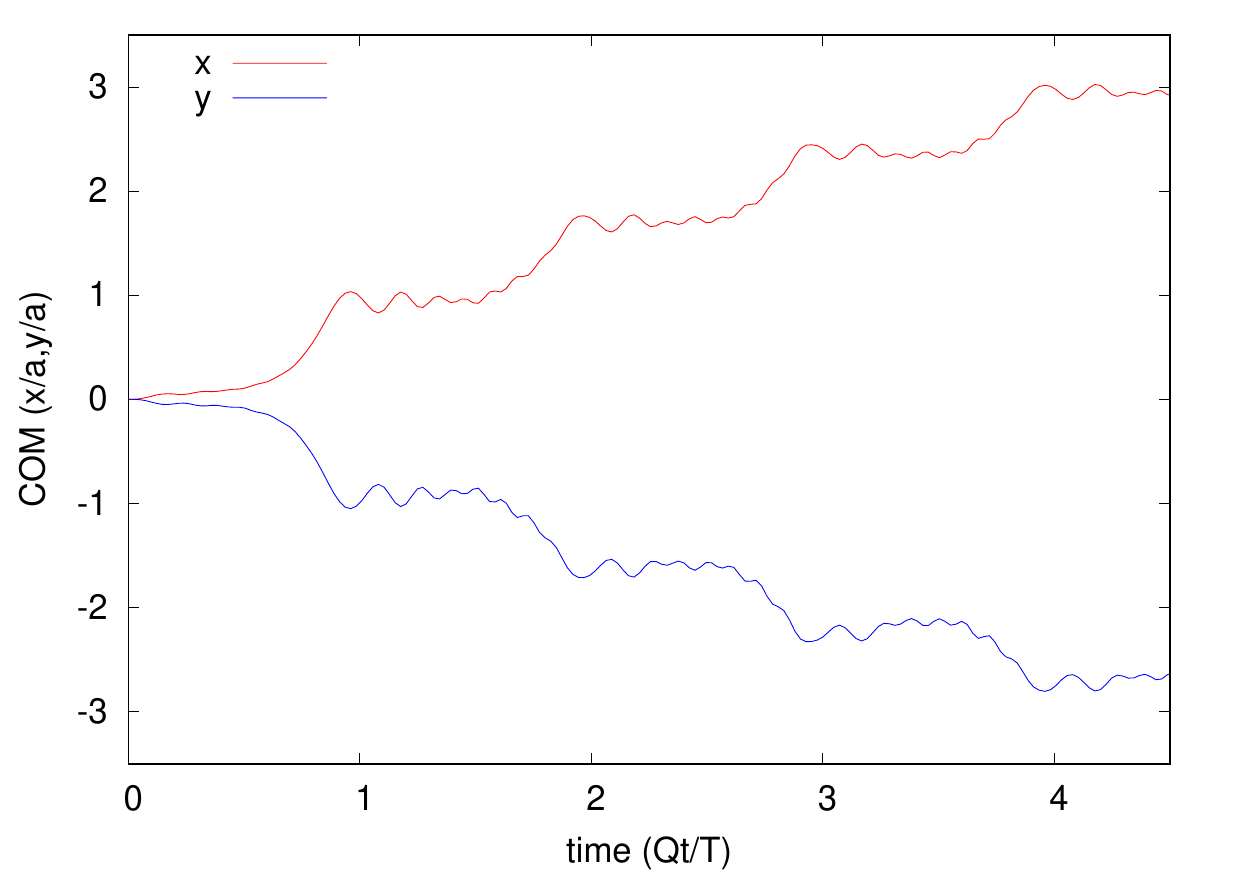} \\
(b) \\ \includegraphics[width=7cm]{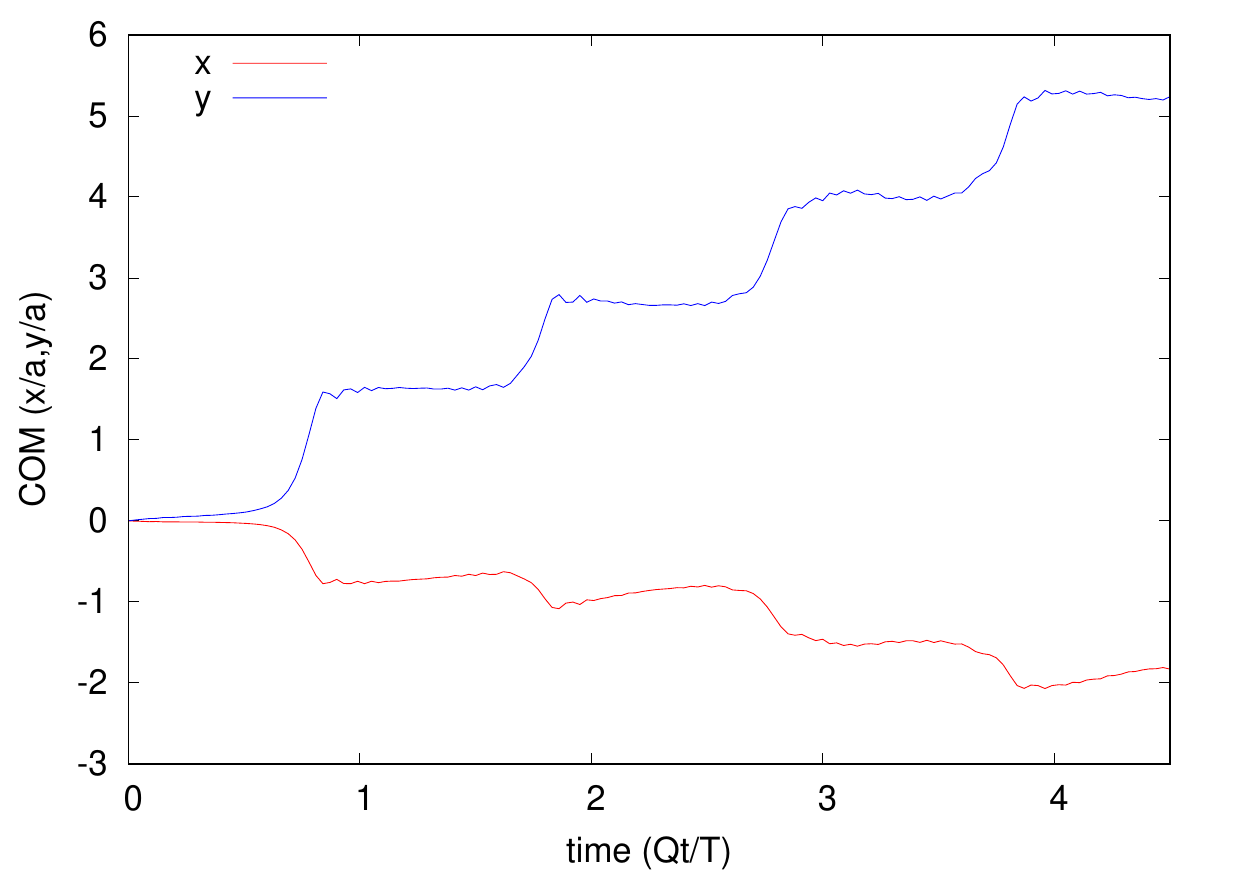}
\caption{(Color online) Time evolution of the average COM for the five lowest energy states in the case of $V_{\mathrm{L}}=1.5$, $V_{\mathrm{SL}}=0.5, V_{\mathrm{trap}}=0.00001$ and (a) $q_x = 3, q_y = 4$ and (b) $q_x = 3, q_y = 5$. }
\label{34_35}
\end{center}
\end{figure}

First, we consider the region near the tight-binding limit, i.e. $V_{\mathrm{L}} \gg V_{\mathrm{SL}}^2/(4E_{\mathrm{R}})$, where $E_\mathrm{R}$ denotes the recoil energy \cite{RevModPhys.80.885}. 
In the previous discussions, we calculated the amount of pumping by considering the integral of the Berry curvature, but this corresponds to the amount of pumping when the perfectly adiabatic process is achieved. In the present case with a trap potential, we must solve the time-dependent Schr\"{o}dinger equation represented by Eq. (\ref{timedepham}). We need a new parameter of one cycle time $T$.  In practice, the perfectly adiabatic process cannot be realized.
Although the quantization of the amount of pumping is lost due to the effect of the harmonic trap and the loss of the adiabaticity, the correct tendency remains for the direction and the amount of pumping. 

Figure \ref{harmeff} shows the COM for the five states with the lowest energy for various time periods $T$. It shows that when $T$ is chosen so that the pumping is performed at appropriate speeds, the COM after one cycle is around $(u,v)=(1,-1)$. We can see that plateaus are found in the middle of the figure. In other words, the solutions to the Diophantine equation, discussed in the previous section, appear as the amount of pumping only when $T$ is in the range of intermediate speeds. Here, we discuss why the plateaus appear in Fig. \ref{harmeff}. There are three important time scales, the period of the cycle $T$, the inverse of the band-gap size $2\pi/E_{\mathrm{gap}}$, and the inverse of the depth of the harmonic trap $2\pi/E_{\mathrm{trap}}$. When $T$ is much smaller than $2\pi/E_{\mathrm{gap}}$, the pumping procedure is too fast and cannot be regarded as an adiabatic process. On the other hand, when $T$ is much larger than $2\pi/E_{\mathrm{trap}}$, there is enough time for the COM of particles to return to the initial position, that is, the center of the harmonic trap. As a result, $T$ should be between $2\pi/E_{\mathrm{gap}}$ and $2\pi/E_{\mathrm{trap}}$ to observe clear pumping behavior. In other words, if $T$ is too small, the particles will not be able to follow the motion of the superlattice and will be left behind, while if $T$ is too large, because of the harmonic trap effect, the pumping does not work properly.

Figure \ref{34_35} shows concrete examples of such pumping. Figures \ref{34_35}(a) and \ref{34_35}(b) show the time evolutions of the COM for the five lowest-energy states in the case of $q_x = 3, q_y = 4$ and $q_x = 3, q_y = 5$, respectively. According to Fig. \ref{34_35}, particles are pumped in the $(1, -1)$ direction for (a) and the $(-1, 2)$ direction for (b). Surprisingly, while the direction of the superlattice is changed only slightly, the pumping direction is completely changed. This pumping direction can be predicted by solving Eq. (\ref{3Ddiophantos}) with the condition $r=1$ since the contribution of the lowest band is dominant. For example, in case (a), the solutions of Eq. (\ref{3Ddiophantos}) are $(u, v) = (1+3m, -1+4n)$ with integers $m$ and $n$. $|u|$ and $|v|$ correspond to the order of perturbation. Therefore, although there are an infinite number of solutions, only the solution where $|u|$ and $|v|$ have the smallest values is dominant. In case (a), this solution is $(t_x, t_y) = (1, -1)$. Similarly, in case (b), the solution of the Diophantine equation is $(u, v) = (-1+3m, 2+5n)$, and the dominant solution is $(u, v) = (-1, 2)$. These results are consistent with Fig. \ref{34_35}.

Observing Fig. \ref{34_35} carefully, we find a small oscillation after the particle was pumped to the next lattice minimum. This oscillation occurs at the moment when the ground-state energies at two sites become close. At that moment, the potential shape can be approximated by a double well, and the particles oscillate between the two wells. Since the angular frequency of this oscillation will be the same as the hopping amplitude $J$, the period will be $T_{\mathrm{DW}} = 2\pi / J$. To estimate $J$ in the optical lattice, we can use $J = \frac{4E_{\mathrm{R}}}{\sqrt{\pi}} s^{3/4} \exp (-2\sqrt{s})$, where $s = V_\mathrm{L}/E_\mathrm{R}$.  We have checked that this estimate is consistent with the numerical result in Fig. \ref{34_35}. We have also verified that the frequency of this small oscillation does not change when we change the time period $T$. This result implies that the small oscillation is caused by the pumping, which is performed in a finite time. 

\begin{figure}[t]
\centering \includegraphics[width=7cm]{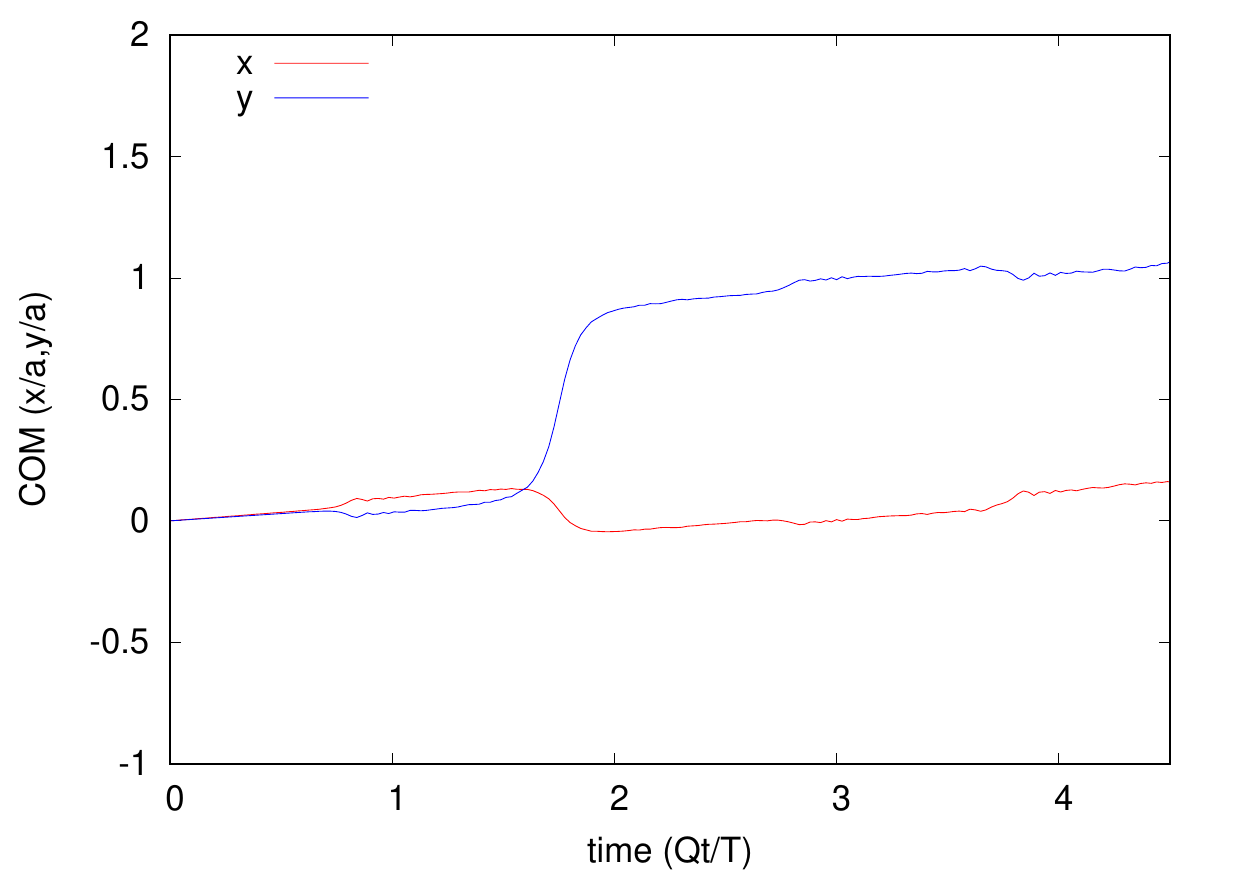}
\caption{(Color online) Time evolution of the average COM for the five lowest energy states in the case of $V_{\mathrm{L}}=1.5$, $V_{\mathrm{SL}}=5.0, V_{\mathrm{trap}}=0.00001, q_x = 3, q_y = 4$.}
\label{05to5}
\end{figure}

\begin{figure*}[t]
(a)\\ \includegraphics[width=16cm]{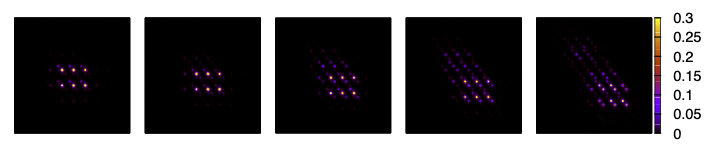}\\
(b)\\ \includegraphics[width=16cm]{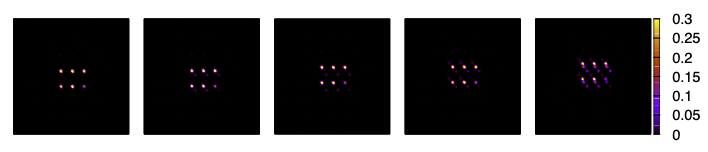}\\
{\sf \hspace{-0.95cm} 0 \hspace{2.4cm} 1.2 \hspace{2.35cm} 2.4 \hspace{2.35cm} 3.6 \hspace{2.35cm} 4.8\\
\hspace{-1.2cm} time (Qt/T)}
\caption{(Color online) Time dependence of particle density profiles for the five lowest energy states in the case of $q_x = 3, q_y = 4$, $V_{\mathrm{L}}=1.5$, with (a) $V_{\mathrm{SL}} = 0.5$ and (b) $V_{\mathrm{SL}} = 5.0$. (a) is an example of the density profiles in the Hofstadter region and (b) is an example of those in the rectifying region. Horizontal and vertical axes correspond to the $x$- and $y$-directions, respectively. In (a), particles move in the positive $x$ (right) and negative $y$ (down) direction. On the contrary, in (b), they move in the positive $y$ (up) direction.  
}
\label{05to5density}
\end{figure*}

Throughout this paper, we have focused on the case of $p_x=p_y=1$, but the Diophantine equation (\ref{3Ddiophantos}) can also be applied to other cases. For the case where $p_x\ne 1$ and $p_y\ne 1$, see Appendix \ref{pxpynot1}.

\subsubsection{Rectifying Region}

In the opposite region, where the $V_\mathrm{SL}$ term is dominant, the pumping behavior changes completely. From our numerical simulation, we find that the pumping direction is restricted to the exactly $x$-axis or the $y$-axis in this region. More precisely, the pumping direction will be the $x$-axis or the $y$-axis, whichever is closer.

When $V_\mathrm{SL}$ varies across the Hofstadter region to the rectifying region, a crossover occurs. Figures \ref{34_35}(a), \ref{05to5}, and \ref{05to5density} provide concrete examples of such a crossover. Figures \ref{34_35}(a) and \ref{05to5} show the time evolution of the COM for the five lowest energy states in the case of $q_x = 3, q_y = 4$, with (a) $V_\mathrm{SL} = 0.5$ and (b) $V_\mathrm{SL} = 5.0$. We note that only the difference in the superlattice potential depth causes the change in pumping direction. Figures \ref{05to5density}(a) and \ref{05to5density}(b) show the time dependence of particle density profiles computed under the conditions corresponding to Fig. \ref{34_35}(a) and Fig. \ref{05to5}, respectively. We can clearly see that particles move in different directions depending on whether $V_\mathrm{SL}$ is small or large.

The band structure in the previous section gives us a hint to understand why this crossover happens. The solution of the Diophantine equation with $r=1$ appears when only the lowest band is considered, and pumping occurs in both the $x$- and $y$-directions. On the contrary, if we compute a state that is lower than the wide band gap,  pumping will occur only in the $x$- or $y$-direction. The result in Fig. \ref{05to5} suggests that the state below this wide band gap is occupied in this region. A system containing a harmonic trap has the property that a wide band gap is automatically selected without adjusting the number of particles. In the present case, by changing $V_{\mathrm{SL}}$, the width of the band gap also changes, and the crossover will be caused by the change in the band gap, which is automatically selected by the effect of the harmonic trap.

We note that such a change in the direction of pumping has also been studied by Lohse {\it et al.}\cite{NaturePhys.12.350}.
In Ref. \onlinecite{NaturePhys.12.350}, the topological transition in 1D systems is reported, where the Chern number of excited states is changed. On the other hand, the crossover found in our study is due to the effect of the harmonic trap. Since the coefficients in the Bloch wave expansions of the trapped eigenstates evolve continuously as functions of $V_\mathrm{SL}$ and are not quantized because of the effect of the harmonic trap, quantization of the pumping is incomplete and the topological transition does not occur.

\section{Conclusion\label{conclusion}}
We have proposed a 2D version of Thouless pumping by employing a 2D square lattice tight-binding model with an obliquely introduced superlattice. We have shown that quantized particle transport occurs in this system, and the transport is expressed as a solution of a Diophantine equation. This topological nature can be understood by mapping the Hamiltonian to a 3D cubic lattice model with a homogeneous magnetic field. Moreover, we have investigated how the harmonic trap affects our model by solving the time-dependent Schr\"{o}dinger equation. We have shown that plateaus appear in the plot of the COM against the time period $T$ for superlattice sliding. When $T$ is in the plateau region, nearly quantized pumping is achieved.
We have found that the pumping behavior is different in two regions, which we call the Hofstadter region and the rectifying region. 
In the Hofstadter region, the amount of pumping obeys the Diophantine equation. On the contrary, in the rectifying region, the pumping direction is restricted to exactly the $x$-axis or the $y$-axis direction. This difference causes the crossover behavior,  highlighting the effect of the harmonic trap. 

In summary, we have shown that the amount of pumping in a 2D lattice with an obliquely introduced superlattice obeys a Diophantine equation, and the harmonic trap affects the pumping behavior in various ways. The effects of the interaction and the lattice deformation are left for future studies.

\begin{acknowledgments}
This work was supported by JSPS KAKENHI (Grant No. JP18H01140, JP19H01838 (N. K.), Grant No. JP17K17822 and JP20H05270 (M. T.)) and a Grant-in-Aid for Scientific Research on Innovative Areas ``Topological Materials Science" (Grant No. JP15H05855).
\end{acknowledgments}

\appendix

\section{Derivation of Diophantine Equation in Tight-Binding Limit}
\label{diotight}
In this Appendix, we derive Eq. (\ref{3Ddiophantos}) from the Hamiltonian
\begin{equation}
\begin{aligned}
{\cal H}(\phi)&=t_x\sum_{m,n} \left[ \hat{c}^{\dag}_{m+1,n}(\phi)\hat{c}_{m,n}(\phi)+\mathrm{h.c.} \right]  \\ &+t_y\sum_{m,n} \left[ \hat{c}^{\dag}_{m,n+1} (\phi )\hat{c}_{m,n}( \phi) +\mathrm{h.c.} \right]  \\
 &+ V \sum_{m,n} \cos \left( 2\pi \left( \Phi_x m+\Phi_y n \right) +\phi \right)\hat{c}^{\dag}_{m,n}(\phi)\hat{c}_{m,n}( \phi ).
\end{aligned}
\end{equation}
This Hamiltonian is periodic in $\phi$ with period $2\pi$. Here, we suppose that $\Phi_x$ and $\Phi_y$ are rational numbers, such as $\Phi_x = p_x/q_x, \Phi_y = p_y/q_y$, where $(p_x, q_x)$ and $(p_y, q_y)$ are pairs of coprime integers with $q_x, q_y > 0$.

The Fourier-transformed Hamiltonian obtained by using Eq. (\ref{hamiltonian3D}) is
\begin{equation}
\begin{aligned}
H &= - \int _ { - \pi } ^ { \pi } \frac { d k _ { x } } { 2 \pi } \int _ { - \pi } ^ { \pi } \frac { d k _ { y } } { 2 \pi } \\ & \biggl[ t \left( \cos \left( k _ { x } \right) + \cos \left( k _ { y } \right) \right) \hat{c} ^ { \dagger } ( k _ { x } , k _ { y } , k _ { z } )  \hat{c} ( k _ { x } , k _ { y }, k _ { z } ) \\  & + \frac{V}{2} \Bigl( e ^ { - i k _ { z } } \hat{c} ^ { \dagger } ( k _ { x } + 2 \pi \Phi_x , k _ { y } + 2 \pi \Phi_y, k_z ) \hat{c} ( k _ { x } , k _ { y }, k_z ) \\  &+ e ^ { i k _ { z } } \hat{c} ^ { \dagger } ( k _ { x } - 2 \pi \Phi_x , k _ { y } - 2 \pi \Phi_y, k_z ) \hat{c} ( k _ { x } , k _ { y }, k_z ) \Bigl) \biggr] . 
\end{aligned}
\label{Aeq2}
\end{equation}
Here, we have defined $k_z = \phi$ for convenience. However, there is mixing between $(k_x, k_y, k_z)$ and $(k_x\pm 2 \pi \Phi_x, k_y \pm 2 \pi \Phi_y, k_z )$, and thus the Hamiltonian is not diagonal in $\bf{k}$. To diagonalize it, we must separate $k_x, k_y$ into $q_x \times q_y$ regions as follows:
\begin{align}
k_x & = k_x^0 + 2 \pi \Phi_x m , \\
k_y & = k_y^0 + 2 \pi \Phi_y n .
\end{align}
Then, the Hamiltonian can be written as
\begin{equation}
H = \frac { 1 } { ( 2 \pi ) ^ { 3 } } \int _ { - \pi / q_x } ^ { \pi / q_x } d k _ { x } ^ { 0 } \int _ { - \pi / q_y } ^ { \pi / q_y } d k _ { y } ^ { 0 } \int _ { - \pi } ^ { \pi } d k _ { z } \hat { H } ( k _ { x } ^ { 0 } , k _ { y } ^ { 0 } , k _ { z } ),
\end{equation}
\begin{widetext}
\begin{equation}
\begin{aligned} \hat { H } ( k _ { x } ^ { 0 } , k _ { y }^ { 0 }, k_z ) &=  \sum _ { m = 0 } ^ { q_x - 1 } \sum _ { n = 0 } ^ { q_x - 1 }    \left\{ - 2 t \left( \cos \left( k _ { x } ^ { 0 } + 2 \pi \Phi_x m \right) + \cos \left( k _ { y } ^ { 0 } + 2 \pi \Phi_y n \right) \right) \right. \\ & \left. \times \hat{c}^ { \dagger } \left( k _ { x } ^ { 0 } + 2 \pi \Phi_x m , k _ { y } ^ { 0 } + 2 \pi \Phi_y n , k _ { z } \right) \hat{c} \left( k _ { x } ^ { 0 } + 2 \pi \Phi_x m , k _ { y } ^ { 0 } + 2 \pi \Phi_y n , k _ { z } \right) \right. \\ & \left. - \frac{V}{2} \left( e ^ { - i k _ { z } } \hat{c} ^ { \dagger } \left( k _ { x } ^ { 0 } + 2 \pi \Phi_x ( m + 1 ) ,  k _ { y } ^ { 0 } + 2 \pi \Phi_y (n + 1) , k _ { z } \right) \hat{c} \left(  k _ { x } ^ { 0 } + 2 \pi \Phi_x m , k _ { y } ^ { 0 } + 2 \pi \Phi_y n , k _ { z } \right)  \right. \right.  \\ & \left. \left. + e ^ { i k _ { z } } \hat{c}^ { \dagger } \left(  k _ { x } ^ { 0 } + 2 \pi \Phi_x ( m - 1 ) ,  k _ { y } ^ { 0 } + 2 \pi \Phi_y ( n - 1 ) , k _ { z }\right)  \hat{c} \left( k _ { x } ^ { 0 } + 2 \pi \Phi_x m , k _ { y } ^ { 0 } + 2 \pi \Phi_y n , k _ { z }  \right) \right) \right\} . \end{aligned}
\end{equation}
\end{widetext}
Note that the Brillouin zone is reduced to $[-\pi /q_x, \pi /q_x]$ for $k_x$ and $[-\pi /q_y, \pi /q_y]$ for $k_y$.
Now, the Schr\"odinger equation $\hat { H } \left( k _ { x } ^ { 0 } , k _ { y } \right) | \psi \rangle = E _ { k _ { x } ^ { 0 } , k _ { y } } | \psi \rangle$ is reduced to that for a 1D tight-binding model. The single-particle energy is obtained by expanding the state into single-particle states at each lattice point $m$,
\begin{equation}
| \psi \rangle = \sum _ { m = 0 } ^ { q - 1 } a _ { m } \hat{c} ^ { \dagger } ( k _ { x } ^ { 0 } + 2 \pi \Phi_x m , k _ { y } ^ { 0 } + 2 \pi \Phi_y m , k _ { z } ) | 0 \rangle ,
\end{equation}
where $| 0 \rangle$ is the vacuum. The eigenvalue equation is
\begin{equation}
\begin{aligned}
\left( - 2 t  \cos \left( k _ { x } ^ { 0 } + 2 \pi \Phi_x n \right) - 2 t  \cos \left( k _ { y } ^ { 0 } + 2 \pi \Phi_y n \right) \right) a _ { n } \\ - \frac{V}{2} \left( e ^ { - i k _ { z } } a _ { n - 1 } + e ^ { i k _ { z } } a _ { n + 1 } \right) = E _ { k _ { x } ^ { 0 } ,k _ { y } ^ { 0 } , k _ { z } } a _ { n } .
\label{3Dharper_app}
\end{aligned}
\end{equation}
This corresponds to the Harper equation for the Hofstadter model. Let $Q$ be the least common multiple of $q_x$ and $q_y$, and let $N_x$ and $N_y$ be integers that satisfy $N_x/Q=p_x/q_x=\Phi_x$, $N_y/Q=p_y/q_y=\Phi_y$. For convenience, we perform the transformation
\begin{equation}
a _ { j } = \sum _ { l = 0 } ^ { Q-1 } e ^ { i 2 \pi j l/Q } b _ { l } ,
\end{equation}
and we obtain 
\begin{equation}
\begin{aligned}
&-t_x\left( e^{ik_x^0} b_{j+N_x} + e^{-ik_x^0} b_{j-N_x}\right) \\ &-t_y \left( e^{ik_y^0} b_{j+N_y} + e^{-ik_y^0} b_{j-N_y} \right) \\ &-V \cos \left( k_z + \frac{2\pi j}{Q}\right) b_j = E_{k_x^0, k_y^0, k_z} b_j .
\end{aligned}
\end{equation}
To solve this equation, we apply the perturbation theory under the condition $t_x, t_y \ll V$. The solution at $t_x, t_y = 0$ is
\begin{equation}
E _ { m } \left( k _ { x } ^ { 0 } , k _ { y } ^ { 0 } , k_z \right) = - 2 t _ { a } \cos \left( k _ { z } + \frac{2 \pi m}{Q} \right) , \quad \psi _ { j } = \delta _ { j , m } .
\end{equation}
If we make $t_x, t_y$ finite, gaps open at the degeneracy points. To understand this in detail, we should make the condition for degeneracy clear. When the two bands $\psi_{m_1}$ and $\psi_{m_2}$ cross each other, the degeneracy condition is 
\begin{equation}
k _ z + \frac { 2 \pi } { Q } m _ { 1 } = - \left( k _ z+ \frac { 2 \pi } { Q } m _ { 2 } \right) \quad (\bmod \ 2 \pi) .
\end{equation}
The degeneracy only occurs when $k_z = 0, \pm \pi / q$, so we can rewrite it as
\begin{equation}
m_1+m_2+l \equiv 0 \quad (\bmod \ q) \ \ (l\in \{ 0, \pm 1 \}, l=q k_x^0/\pi) .
\label{banddege}
\end{equation}
It is obvious that the lowest band corresponds to $m=0$. It is possible to determine all of the band indices by using Eq. (\ref{banddege}). There is a simple relation between the gap number $r$ and band indices $m_1, m_2$:
\begin{equation}
r \equiv - |m_1-m_2| \quad (\bmod \ Q) .
\end{equation}
Now we are ready to use the perturbation theory. Since the $t_x(t_y)$ term only mixes two sites that are $P_x(P_y)$ sites apart from each other, in order to hybridize $\phi_{m_1}$ and $\phi_{m_2}$, the following equation must be satisfied:
\begin{equation}
|m_1-m_2| = N_x t_r + N_y u_r \quad (\bmod \ Q),
\end{equation}
that is, 
\begin{equation}
\frac{r}{Q} = s_r + \frac{p_x}{q_x} u_r + \frac{p_y}{q_y} v_r .
\label{3Ddio}
\end{equation}
The lowest order of perturbation that mixes $\phi_{m_1}$ and $\phi_{m_2}$ is the $|u_r|$th order for the $t_x$ term and the $|v_r|$th order for the $t_y$ term, respectively. 

The Hamiltonian around the $r$th gap is
\begin{equation}
\left( \begin{array} { cc }  \epsilon & { \Delta e ^ {- i k _ { x } u_ { r } }  e ^ {- i k _ { y } v _ { r } } }  \\ { \Delta e ^ { i k _ { x } u_ { r }  }e ^ { i k _ { y } v_ { r } } }  & { - \epsilon } \end{array} \right) \left( \begin{array} { l } { a } \\ { b } \end{array} \right) = E \left( \begin{array} { l } { a } \\ { b } \end{array} \right).
\end{equation}
Now, it is possible to calculate the amount of pumping in one cycle from this Hamiltonian. The amount of pumping will be $t_r$ for the $x$-direction and $u_r$ for the $y$-direction, and this is exactly the solution of the Diophantine equation (\ref{3Ddio}). However, we should be aware that ``one cycle" here corresponds to the width of the Brilliouin zone of $k_z$, while $k_z$ changes its value by $2\pi/Q$.

\section{Justification of Diophantine Equation in Plane-wave Approximation}
\label{dioplane}
In this Appendix, we derive Eq. (\ref{3Ddiophantos}) from the Hamiltonian given in Eq. (\ref{continuum1}). 
For simplicity, let $q_x=q_y=1$. To make the perturbation theory easier to apply, we divide the Hamiltonian in Eq. (\ref{continuum1}) into $H_0$ and $H_1$,
\begin{eqnarray}
\label{continuum3}
H &&= H_0 + H_1, \\
H_0 &&= \int dx dy \psi^\dagger (x,y) \biggl[ -\cfrac{\hbar^2}{2m} \cfrac{\partial^2}{\partial x^2} -\cfrac{\hbar^2}{2m}\cfrac{\partial^2}{\partial y^2} \nonumber \\
&&+ V_x(x,y) + V_y(x,y) \biggl] \psi(x,y), \\
V_j(x,y) &&= 2v_j \cos(q_j j) \ \ \ \ (j = x,y), \\
H_1 &&= \int dx dy \psi^\dagger (x,y)V_{\mathrm{SL}}(x,y)\psi(x,y), \\
V_{\mathrm{SL}}(x,y) &&= 2v_{\mathrm{SL}} \cos(p_x x + p_y y - \varphi).
\label{continuum4}
\end{eqnarray}

First, eigenstates of $H_0$ can be written in Bloch states in a Brillouin zone of size $(2\pi/a,2\pi/a)$. If $H_1$ is added here, some Bloch states are hybridized. This is an idea just like the magnetic Brillouin zone.
Let the Bloch state obtained from $H_0$ be $\hat{\psi}(k_x, k_y)$ and the corresponding eigenenergies be $E_0(k_x, k_y)$. The expression corresponding to Eq. (\ref{Aeq2}) in Appendix A is
\begin{equation}
\begin{aligned}
H &= - \int _ { - \pi } ^ { \pi } \frac { d k _ { x } } { 2 \pi } \int _ { - \pi } ^ { \pi } \frac { d k _ { y } } { 2 \pi } \\ & \biggl[ E_0(k_x, k_y) \hat{\psi} ^ { \dagger } ( k _ { x } , k _ { y } , k _ { z } )  \hat{\psi} ( k _ { x } , k _ { y }, k _ { z } ) \\  & + \frac{V}{2} \Bigl( e ^ { - i k _ { z } } \hat{\psi} ^ { \dagger } ( k _ { x } + 2 \pi \Phi_x , k _ { y } + 2 \pi \Phi_y, k_z ) \hat{\psi} ( k _ { x } , k _ { y }, k_z ) \\  &+ e ^ { i k _ { z } } \hat{\psi} ^ { \dagger } ( k _ { x } - 2 \pi \Phi_x , k _ { y } - 2 \pi \Phi_y, k_z ) \hat{\psi} ( k _ { x } , k _ { y }, k_z ) \Bigl) \biggr] . 
\end{aligned}
\end{equation}
This is not diagonal because it is mixed for different $(k_x, k_y)$. However, we can diagonalize it by using a similar method to that in Appendix A. 
By expressing the single-particle states as a linear combination of different wave number states,
\begin{equation}
| \psi \rangle = \sum _ { m = 0 } ^ { q - 1 } a _ { m } \hat{\psi} ^ { \dagger } ( k _ { x } ^ { 0 } + 2 \pi \Phi_x m , k _ { y } ^ { 0 } + 2 \pi \Phi_y m , k _ { z } ) | 0 \rangle, 
\end{equation}
then the eigenvalue equation is obtained as follows:
\begin{equation}
\begin{aligned}
\left( - 2 t \ \psi \left( k _ { x } ^ { 0 } + 2 \pi \Phi_x n \right) - 2 t \ \psi \left( k _ { y } ^ { 0 } + 2 \pi \Phi_y n \right) \right) a _ { n } \\ - \frac{V}{2} \left( e ^ { - i k _ { z } } a _ { n - 1 } + e ^ { i k _ { z } } a _ { n + 1 } \right) = E _ { k _ { x } ^ { 0 } ,k _ { y } ^ { 0 } , k _ { z } } a _ { n } .
\label{3Dharper}
\end{aligned}
\end{equation}

Since the eigenenergies of the Bloch state do not change so much from the cosine function, the position of the degenerate point does not change. As a result, the degree of perturbation required is determined from the degenerate point condition, which leads to the same Diophantine equation.

\section{Pumping in the $p_x \ne 1, p_y \ne 1$ Case}
\label{pxpynot1}

\begin{figure}[t]
\centering \includegraphics[width=7cm]{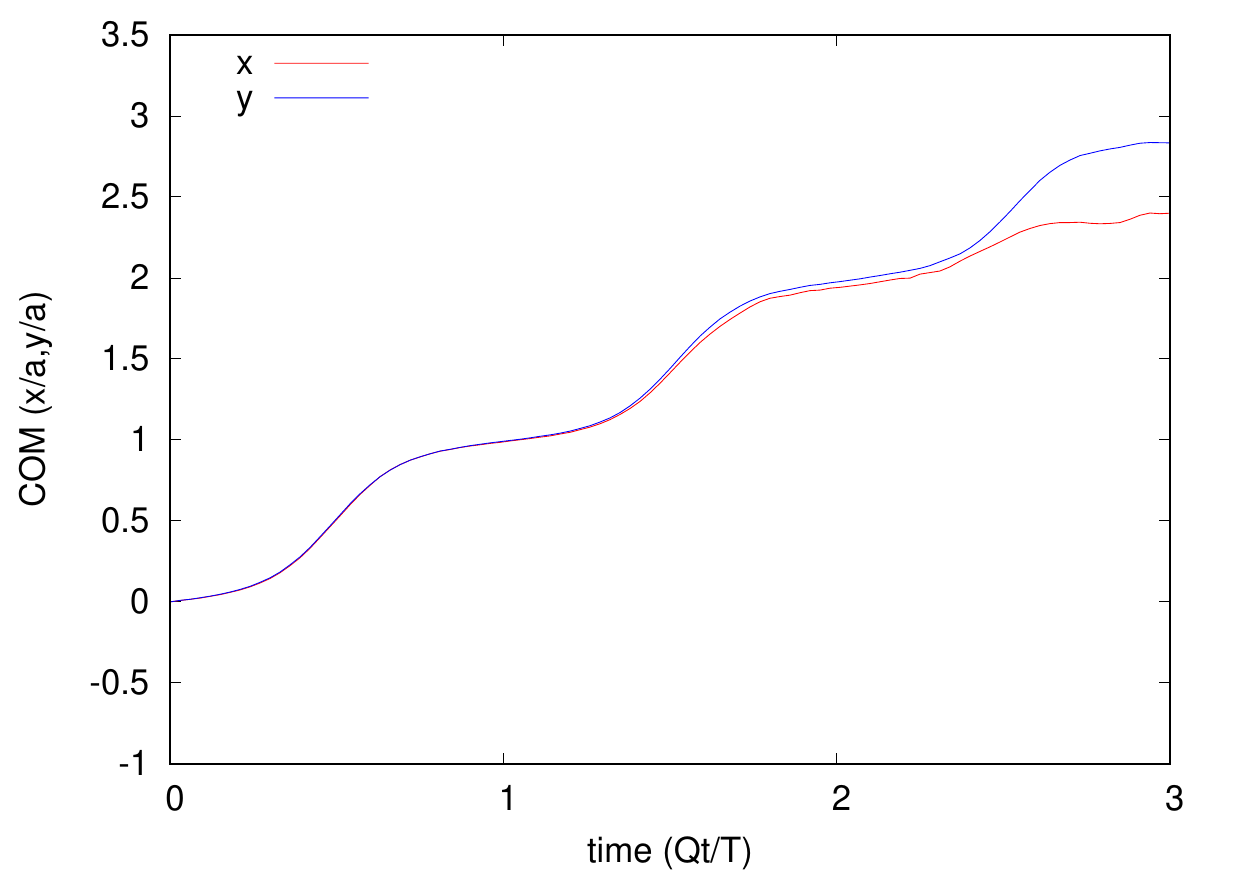}
\caption{(Color online) Time evolution of the average COM for the five lowest energy states in the case of $V_{\mathrm{SL}}=5.0, V_{\mathrm{trap}}=0.00001, p_x/q_x = 2/3, p_y/q_y = 2/5$. }
\label{235}
\end{figure}

Throughout this paper, we have focused on the $p_x=p_y=1$ case. However, the obtained Diophantine equation (\ref{3Ddiophantos}) is appliacable to other cases. In this Appendix, we show a concrete example for these cases. 

Figure \ref{235} shows the time evolution of the average COM in the case of $p_x/q_x = 2/3, p_y/q_y = 2/5$ where the harmonic potential exists. Since $1\cdot 2/3+1\cdot 2/5=16/15=1+1/15$, $(u,v)=(1,1)$ is the solution of the Diophantine equation that corresponds to the lowest order of the perturbation. If we look at Fig. \ref{235}, we see that particles are pumped in the direction of the line $x=y$. We can confirm that the pumping direction and the solution of the Diophantine equation also match each other in this case.

\bibliographystyle{jpsj}
\bibliography{69803}

\begin{thebibliography}{10}
\bibitem{inverse}
In our model, the inversion symmetry is restored when one of the sites is on
  the minimum line of the potential. Since there are $q_x q_y$ sites in the
  superlattice produced by the on-site potential, there are $q_x q_y$ points
  protected by inversion symmetry in one cycle in our model.
\end{thebibliography}

\end{document}